# Tamm plasmon polariton in planar structures: A brief overview and applications

Chinmaya Kar[1], Shuvendu Jena[*,2], Dinesh V. Udupa[2], K. Divakar Rao[1]

[1]Photonics & Nanotechnology Section, Atomic & Molecular Physics Division, Bhabha Atomic Research Centre Facility, Visakhapatnam 531011, India

[2]Atomic & Molecular Physics Division, Bhabha Atomic Research Centre, Mumbai 400085, India

* Corresponding author
*E-mail addresses*: shuvendujena9@gmail.com, shujena@barc.gov.in (S. Jena)

**Abstract:** Tamm plasmon provides a new avenue in plasmonics of interface states in planar multilayer structures due to its strong light matter interaction. This article reviews the research and development in Tamm plasmon polariton excited at the interface of a metal and a distributed Bragg reflector. Tamm plasmon offers an easy planar solution compared to patterned surface plasmon devices with huge field enhancement at the interface and does not require of any phase matching method for its excitation. The ease of depositing multilayer thin film stacks, direct optical excitation, and high-Q modes make Tamm plasmons an attractive field of research with potential practical applications. The basic properties of the Tamm plasmon modes including its dispersion, effect of different plasmon active metals, coupling with other resonant modes and their polarisation splitting, and tunability of Tamm plasmon coupled hybrid modes under externally applied stimuli have been discussed. The application of Tamm plasmon modes in lasers, hot electron photodetectors, perfect absorbers, thermal emitters, light emitting devices, and sensors have also been discussed in detail. This review covers all the major advancements in this field over the last fifteen years with special emphasis on the application part.

# 1. Introduction

Surface electromagnetic waves are the electromagnetic modes excited at the interface of two different materials, which decays evanescently from the interface. During last few decades, many surface electromagnetic states have been discovered and investigated for understanding their physics. These include modes due to Zenneck wave, Dyakonov state, surface plasmon wave, optical Tamm state, and topological edge state [1-3]. Zenneck wave is simply a plane wave solution to Maxwell's equations polarized perpendicular to the interface that separates free space from a half space with a finite conductivity. The amplitude of this wave decays exponentially in the directions both parallel and perpendicular to the boundary with differing decay constants [4]. Dyakonov surface state forms at the interface between an isotropic and an uniaxial birefringent medium [5]. In recent times, topological photonic states are gaining research interest which is reviewed in reference [6]. There are also reports of surface states excited between linear and nonlinear media [7, 8]. Among all types of surface states, surface plasmon polariton (SPP) is the most diversely studied surface electromagnetic modes that is excited at the metal-dielectric interface [9]. Fundamentally, SPPs arise due to the coupling of coherent oscillation of conduction electrons on the metal surface with the electromagnetic fields inside the dielectric medium. These modes can only exist in transverse magnetic (TM) polarisation with its wave vector lying outside the light cone of the dielectric. Hence various wave vector matching techniques utilizing elements like prisms, gratings, highly focussed beams through microscopes [10] and integration with conventional photonic elements [11] are required for SPP excitation. The Kretschmann excitation scheme [12] and dispersion curve of SPPs are given in Fig. 1(a) and (b), respectively. This scheme utilizes a glass prism to excite SPPs in the air-metal interface. From the dispersion curve, it can be seen that propagation of light in air cannot excite the surface plasmons as the dispersion curve of the surface plasmons (metal-air and metal-prism interfaces) do not intersect at any point and lie to the right of the air light line in the Fig. 1 (b). This implies that directly incident light in air has insufficient momentum to excite the surface plasmons for any incidence angle. Light propagating in glass medium and incident on the interface has a higher momentum due to the refractive index (~1.5) of glass. This results in a decrease in gradient of the glass prims light line, intersecting the metal-air interface surface plasmon dispersion curve. The metal-air SPP can thus be excited at that particular light frequency incident from the glass prism side. It also should be noted that the SPP at the metal-prism interface cannot be excited as it is outside the glass prism light line. There are different types of SPPs such as gap, spoof, magneto plasmons and Berreman modes that have been reviewed elsewhere [13], which are useful in sensors, filters, absorbers, photovoltaic devices, and many more photonic applications [14-16].

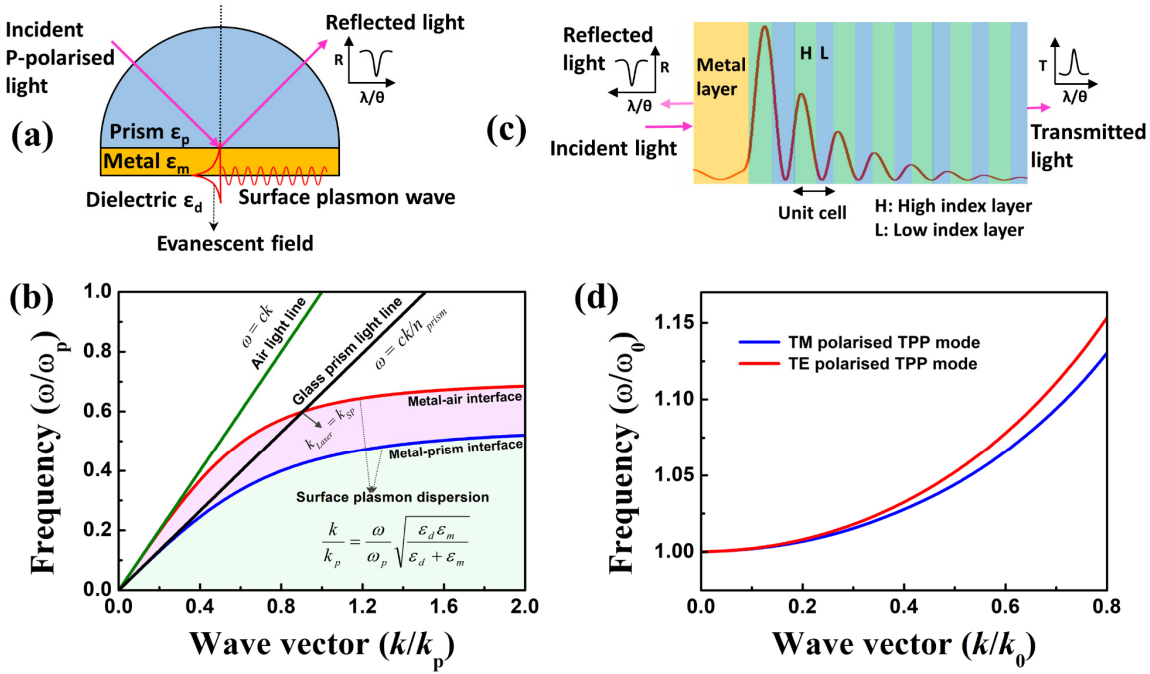

**Fig. 1:** Surface plasmon polariton (SPP) and Tamm plasmon polariton (TPP) configurations with their dispersion curves: (a) Excitation of SPPs at a metal/dielectric interface under Kretschmann geometry. (b) Dispersion curves of SPP for metal-air and metal-prism interfaces along with air and prism light lines. (c) Conventional TPP exciting structure containing a metal and a distributed Bragg reflector (DBR). (d) Dispersion curves of TPP for transverse magnetic (TM) and transverse electric (TE) polarised light.

In recent years, a special type of electromagnetic surface state known as Tamm plasmon polariton (TPP) was theoretically proposed in 2007 [17] and experimentally observed in 2008 [18]. The TPP mode is formed at the interface between a metal with $\varepsilon < 0$ and a one dimensional photonic crystal (1DPC) or a distributed Bragg reflector (DBR) in multilayer planar geometry as shown in Fig. 1(c). It is an optical analog of the so called Tamm electronic states found on atomic layers closest to the surface of a material [19]. The TPP mode can be experimentally observed as a narrow or broad resonance peak or dip in the transmission or reflection spectrum of a sample at wavelengths within the band gap of the 1DPC. The transmission at the Tamm mode becomes perfect when the electric fields in both metal and the DBR are evanescent. Fig. 1(d) shows the dispersion curve of the Tamm plasmon mode and it lies inside the light cone unlike the surface plasmon polariton [17]. It means the TPP modes are polarisation independent, and can be optically excited without any additional phase matching techniques like grating or prism coupling. The TPP mode can be narrowed by using large number of periods in the planar structure and can be used for investigating strong coupling regime. Tamm plasmons exhibit better light matter interactions assisted by strong field confinement at the interface. There has been evidences of Tamm states being transformed into bound states in continuum (BIC) in metal-DBR heterostructures when light is incident at Brewster's angle. Pankin *et al.* [20] experimentally demonstrated the BIC states for the first time by inserting an anisotropic defect layer like liquid crystal (LC) between a 1DPC and a metal. Such structures exhibit Friedrich-Wintgen BICs [21] due to destructive interference of TE and TM polarisations. Wu *et al.* [22] have demonstrated that the Q-factor of quasi-BIC can be tuned by the LC optical axis rotation in a wide range and the quasi-BIC resonance is extremely sensitive to the temperature due to the narrow nematic temperature range of the LC. Bikbaev *et al.* [23] have shown the transformation of Tamm state into BIC under certain conditions in all dielectric nanostructure comprising Si grating on a DBR. The transformation of TPP to BIC has open up a new avenue of research in the field of nanophotonics.

In the early days of its discovery, TPPs attracted attention of research community especially for controlling and enhancement of spontaneous emissions by using microcavity, quantum dots, quantum wells (QWs), and 2D materials [24-26]. These studies have shown modulation of spontaneous emission, strong coupling, hybrid plasmon and exciton coupling with QWs and 2D materials. The ease of depositing the multilayer stacks, direct optical excitation using both TE and TM light, and high-Q modes give Tamm plasmons a potential for interesting practical applications [27]. This review report aims at providing a brief account of TPPs starting from its origin with a small theoretical discussion to explain its characteristic features. We have also discussed its coupling to different fundamental excitations, tunability, polarization splitting, and finally provided an overview of its potential applications.

## 2. Tamm plasmon polariton

Propagation of electrons in crystalline solids are defined by Bloch wave functions obeying Born-von Karman periodic boundary conditions [28]. The electronic states under perfectly periodic potential of lattice exhibit energy bands where no electronic energies are allowed. But in practice, no absolute periodic systems are available in nature as the periodicity terminates at the surface of all crystals. The solution of Schrodinger's equation for such truncated potentials results in two types of eigen solutions; one possessing the pure Bloch character inside the crystal showing energy bands called bulk electronic states and other leads to surface states confined at crystal surface possessing imaginary wave vector perpendicular to the surface called surface electronic states or Tamm states [29]. These states decay into both the vacuum and the bulk crystal with localization at the crystal surface, and appear in the energy gaps of the bulk electronic states. Interestingly, in the optical domain, optical Tamm states (OTS) arise at the interface between two photonic crystal (PC) hetero-structures in the region of overlapping photonic band gaps [30]. PC consists of materials with periodic variation in dielectric constant, which can shape and mould the flow of light inside the PC [31]. The PC is analogous to electronic semiconductor [32] and it generates a range of forbidden wavelength regions called photonic band gaps [33] like electrons suffer in periodic potential of atomic lattice in a semiconductor. The periodicity of the PC can be in one, two or three dimensions. Among these, 1DPCs are the most frequently used as they are easy to fabricate and design, an example being a DBR [34]. OTSs are confined at the interface between both the periodic structures and the waves die out exponentially perpendicular to the interface. When one of the periodic structures is replaced by a plasmonic material that exhibit negative real permittivity like metals, the localized electromagnetic modes couple to the electrons in the metal forming the TPP mode. It is easily identified as narrow resonances in the reflectance and/or transmittance spectrum. The TPP mode possess parabolic dispersion with an effective mass of order of $10^{-5}$ of a free electron mass and the TE-TM polarisation splitting of the TPPs increases quadratically with the in-plane wave vector [17] as shown in Fig. 1(d). The coupling of light to excite TPPs mainly depends upon the 1DPC parameters, metallic layer properties, and angle of incidence. Critical coupling occurs when the reflection minima becomes exactly zero for an optimized photonic hetero-structure [35]. Tamm plasmons exhibit better light matter interaction assisted by strong field confinement at the interface. So, in the early days of its discovery, TPPs attracted attention of the research community especially for controlling and enhancement of spontaneous emissions by using microcavities [36], metallic disk [37], quantum dots [38], quantum wells [39], and nanostructures [40]. Tamm plasmon based structures have been considered as a feasible architecture for developing room temperature polariton lasers [41]. The polariton lasers are highly monochromatic as well as coherent and are based upon the principle of spontaneous emission from the Bose-Einstein condensates of exciton-polariton systems [42]. The hybrid exciton-polariton system is discussed later in detail.

Practically, there will always be decays associated with the TPP modes. The decays are attributed to non-negligible collisional loss in metals, radiative loss through the DBR due to limited number of periods, and roughness of the interface sustaining the TPPs. In addition, metallic layers are associated with scattering loss due to grain boundaries and defects because of crystalline nature of the metallic film. Different metallic layers offer different rate of losses which plays a key role in the reflectance spectra [43]. Most often, silver is used to excite TPPs for its loss characteristics in visible-NIR region. It has been reported that silver has 4.7 and 84 times higher Q-factor of Tamm plasmon resonance than gold and aluminium in visible region, respectively [44]. Similarly, the radiative loss is decreased by increasing the number of periods in the DBR. Quality factor and lifetime are used to quantify decay of the TPP resonance in a photonic hetero-structure. Lifetime of the TPP is found to exhibit strong polarisation and angular dependence. The life time of TPPs (in the order of 10s of femtoseconds) are several times smaller compared to the lifetime of SPPs [45]. This short lifetime makes TPP a strong candidate for applications in all-optical switches and modulators. Apart from planar multilayer structures, Tamm modes can be confined in various directions by nano-patterning the metallic layers. Due to the patterning the Tamm modes become sensitive to the external stimulus which is difficult to realize in planar structures. Buchev *et al*. [46] have generated 3D confined Tamm modes in a meta-structure containing 550 nm large silver nano-disks in square arrays on a DBR containing 11 pairs of niobium pentoxide and silicon dioxide. Confined Tamm states are more tunable and possess high Q factor compared to conventional planar TP structures due to reduced metallic losses. In such patterned structures, the continuous parabolic dispersion nature of the Tamm modes exhibit energy bands [47]. Hence, patterning the metallic layer provides extra degrees of freedom to control the dispersion as well as the electromagnetic field confinement in three dimensions. Qiao *et al*. [48] have experimentally demonstrated Tamm plasmon topological superlattices (TTS) by using specific double layered metasurfaces on a PC. By controlling the topology of the metasurface they observed coupling between TTSs to generate super-modes lying inside the photonic band gap of the PC. However, their existence require the unit cells to be inversion symmetric with metallic and PC bearing opposite Zak phases [49]. Keene and Durach have proposed hyperbolic Tamm plasmons that exist at the interface between a uniaxial metamaterial and a metal [50]. Rudakova *et al*. [51] have numerically studied the existence chiral OTS at the interface of a chiral material like cholesteric liquid crystal (CLC) and multilayer anisotropic mirror followed by a metasurface. They have established that the use of the metasurface can reduce the required number of layers in the multilayer anisotropic mirror at a certain Q-value. Lin *et al*. [52] have observed chiral TPP at the interface of a CLC and a metasurface acting as a half wave plate for polarization and phase matching. Such chiral Tamm structures are sensitive to the configuration of the metasurface, external temperature and orientation of the LC. In present case, we mainly focus on the TPP generated in planar multilayer structures.

## 3. Dispersion and resonant frequency

In-plane dispersion of the TPP mode can be easily derived by solving Maxwell equations in the metal layer [53]. The tangential components of the electric field at the metal layer interface can be expressed as

$$\begin{pmatrix} E \\ B \end{pmatrix} = E_0 \begin{pmatrix} 1 \\ iqc/\omega \end{pmatrix} \quad \text{for TE polarised light} \tag{1}$$

$$\begin{pmatrix} E \\ B \end{pmatrix} = B_0 \begin{pmatrix} 1 \\ iqc/\omega n_m^2 \end{pmatrix} \quad \text{for TM polarised light} \tag{2}$$

where $q^2 = (\omega n_m/c)^2 + k^2$, $\omega$ is the angular frequency of light, $c$ is speed of light, $n_m$ is the complex refractive index of the metal layer, and $k$ is the wave propagation vector. The transfer matrix for a DBR (made of alternate layers of A and B) can be expressed as

$$M = M_A M_B = \begin{pmatrix} t_{11} & t_{12} \\ t_{21} & t_{22} \end{pmatrix} \tag{3}$$

Now the dispersion of TE and TM localized TPP modes can be derived by matching the field vectors in equation (1) and (2) with the eigen-vectors of the transfer matrix shown in equation (3). The eigen-vectors with eigen-values less than unity $e^{-Qd}$ are chosen as they describe the evanescent wave corresponding to the TPP mode. The derived dispersion equations for the TPP in TE and TM polarisations can be expressed as

$$i\sqrt{n_m^2 + \left(\frac{kc}{\omega}\right)^2} = \frac{e^{-Q^{TE} d} - t_{11}^{TE}}{t_{12}^{TE}} \tag{4}$$

$$\frac{i}{n_m^2}\sqrt{n_m^2 + \left(\frac{kc}{\omega}\right)^2} = \frac{e^{-Q^{TM} d} - t_{11}^{TM}}{t_{12}^{TM}} \tag{5}$$

The equations (3) and (4) are used to calculate the dispersion curves (Fig. 1(d)) of a Tamm plasmon photonic structure for TE and TM polarised light.

The resonant dip in the reflectivity spectrum is the signature of TPP mode in the structure. The condition for TP resonance must satisfy the following condition [54]

$$r_M r_{DBR} e^{2iknd} = 1 \tag{6}$$

This equation can be understood as the conditions for constructive round-trip for light inside the cavity formed by the first layer $A$ sandwiched between the metal Ag and the rest of the DBR as shown in Fig. 2(a). The DBR is made up of 8 pairs of $A$ (high index) and $B$ (low index) layers of refractive index 2.07 and 1.45, respectively. The thicknesses of A, B, and Ag are 140 nm, 200 nm, and 25 nm, respectively. The complex refractive index ($n_m$) of silver layer is calculated using Drude's model [54].

$$n_m^2 = \varepsilon_m = \varepsilon_\infty - \frac{\omega_p^2}{\omega^2 - i\gamma\omega} \tag{7}$$

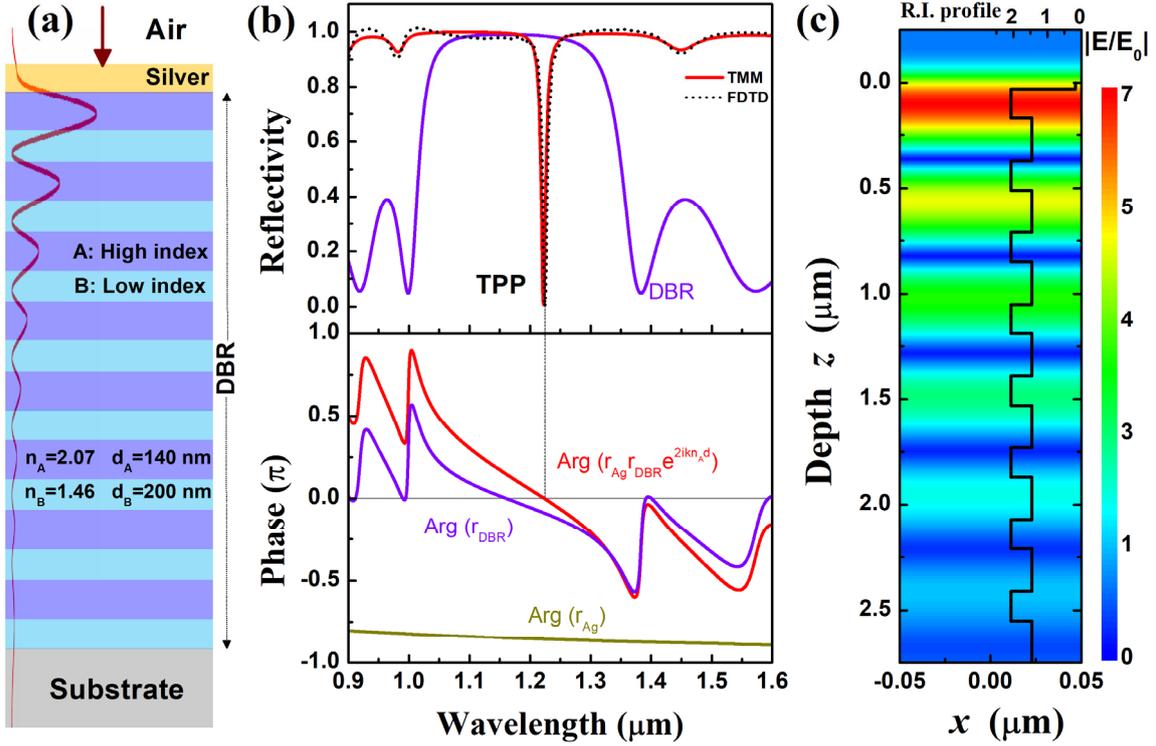

**Fig. 2:** (a) Schematic of metal/DBR structure for excitation of TPP. The top plot of (b) is the reflectivity spectrum of the structure showing Tamm plasmon polariton as a resonant dip obtained using TMM and FDTD method. The bottom plot of (b) presents the argument of reflection co-efficient of Ag, DBR and their combined structure. (c) Electric field intensity profile of the TPP mode obtained using FDTD method along with the refractive index profile (black solid line) of the structure.

where $\varepsilon_\infty$ (=5) and $\omega_p$ (=8.95 eV) are the infinite frequency dielectric constant and plasma frequency of the metal layer, respectively. The reflectivity spectrum of the structures are calculated using transfer matrix method (TMM) [55] and Finite-difference time-domain (FDTD) [56] method. The top panel of Fig. 2(b) shows the high reflection band of the DBR centred at 1160 nm and a sharp resonant reflectivity dip at 1224 nm corresponding to the TP mode. The lower panel of Fig. 2(b) plots the arguments of the reflectivity spectrum for metal, DBR and their combined structure. The phase factor ($e^{2iknd}$) is the phase shift for the cavity round trip in the layer adjacent to the metal layer. The DBR shows a linear phase shift in the band gap region whereas the phase shift due to the metal is nearly constant of about –π. The sum of all phases must be zero *i.e.* $Arg(r_M r_{DBR} e^{2iknd})=0$ at the TP resonance wavelength. This zero-crossing is observed around wavelength of 1224 nm. It exactly matches to the reflectivity minimum of the structure shown in the top panel. It confirms that the resonance observed in the reflectivity spectrum of Fig. 2 (b) is due to the excitation of TPP at the Ag/DBR interface. The approximate value of resonant frequency of a TP mode which lies close to the Bragg frequency ($\omega_0$) can be expressed by the following equation [17]

8 | P a g e

$$\omega_{TP} = \frac{\omega_0}{1 + \left((2n_H \omega_0)/(\beta \omega_p \sqrt{x_0})\right)} \qquad (8)$$

where $x_0 = 1 - (\varepsilon_\infty \omega_0^2 / \omega_p^2)$ [54] $\qquad (9)$

For the structure shown in Fig. 2(a), the value of Bragg energy is $\hbar\omega_0$=1.06 eV; and the observed TP mode has energy $\hbar\omega_{TP}$=1.01 eV. Equation (8) predicts the TP resonant frequency of $\hbar\omega_{TP} \approx 1.02$ eV. The field profile and index profile data for the same photonic hetero-structure has been displayed in Fig. 2(c) which clearly shows that the field is strongly localised in the region near to the metal-DBR interface compared to the entire photonic hetero-structure which is the reason why any perturbation above metallic layers have very small impact on the TP mode. The TP resonance mode depends on the plasma frequency of metal, Bragg frequency, and refractive index contrast of the DBR. These parameters are generally used to tune TPP based devices. In other words, Tamm states are sensitive to any changes in the PC properties like periodicity, refractive index, layer order, insertion of new layers or changing the thickness of the layer adjacent to the metal. The conventional TP structures can be easily modified by introducing spacer layers, quantum wells, quantum dots and mono-layers, but they are difficult to tune over a wide range of wavelengths. As metal layer is very much important for Tamm plasmon excitation, the impact of different plasmon active metals on the properties of TPP mode are discussed in following section.

## 4. Effect of different plasmon active metals

The existence and Q-factor of the Tamm resonance modes depends upon the type of plasmon active metallic layer, thickness of the metallic layer and angle of incidence of light. Several studies on the effect of different metals on the Tamm plasmon modes [57-59] have been reported. To understand the effect of different plasmon active metals on the sensitivity of TPP resonance, a TP structure made of metal/$SiO_2$/DBR has been designed and numerically investigated using TMM. The DBR is made of five periodic bilayers of $TiO_2$ and $SiO_2$ of thicknesses 58 nm and 85 nm, respectively and the thickness of the metallic layer is fixed at 40 nm. The dispersion free refractive index values of $TiO_2$ and $SiO_2$ are assumed to be 2.36 and 1.47, respectively. The refractive index data of the metals Ag, Al, Au, and Cu are taken from the reference [60], while that of Cr is taken from the reference [61]. Fig. 3 (a) and (b) represents the reflectance spectrum of the TP structure for various metals with identical film thickness by injecting light normally from the air-metal and substrate-DBR side, respectively. The figure clearly exhibits a dip within the photonic bandgap which corresponds to the excitation of TPP mode. There is no optical reciprocity in reflectivity spectrum for the TP structures as observed in Fig. 3. This is due to the nonreciprocal electromagnetic field distributions of the TP mode in the structures depending on the incident light direction whether from the metal side or the DBR side [62]. For light incident from metal side (DBR side) in a given TP structure, if the intensity of electromagnetic field localized near the interface between the metal and the DBR is much larger as compared to that of the light incident from the DBR side (metal side), then the absorbance of the structure will be higher at the TP mode wavelength resulting sharper dip in the reflectivity spectrum as compared to that of the DBR side (metal side). It can be observed that no significant TP mode is excited for Cr and Al when light is incident from the metal side where as there is weak TPP excitation when light is incident from DBR side. This happens due to the significant reflection/absorption of the light in the Cr and Al layers resulting poorly localized electric field intensity at the metal/DBR interface for light incident from the metal side. The use of Cr as an active plasmonic layer [63] is reported to exhibit broadband Tamm plasmon polariton. However, the TPP modes are well excited in Ag, Au and Cu irrespective of the direction of incidence. The TPP peaks of Au and Cu almost matches due to similar frequency dependent optical properties of both the metals. Weak impedance matching between the incident light and TPP in the metal/SiO$_2$/DBR geometry is clearly observed form the relatively broad and lesser dip reflectivity spectrum in case of Cr and Al (as compared to Ag, Au, and Cu). The condition for absolute reflectivity minimum requires an exact match of amplitude and phase between the forward and backward propagating beams at the metal/DBR interface [64]. Among all, the Ag based structure exhibits highest Q-factor TP resonance as compared to other metals. Moreover, Ag is primarily chosen for the visible region application owing to its low absorption co-efficient and absence of inter-band transitions in the visible band [65]. It may be noted that the width of the TPP resonance can be controlled by changing the metal layer.

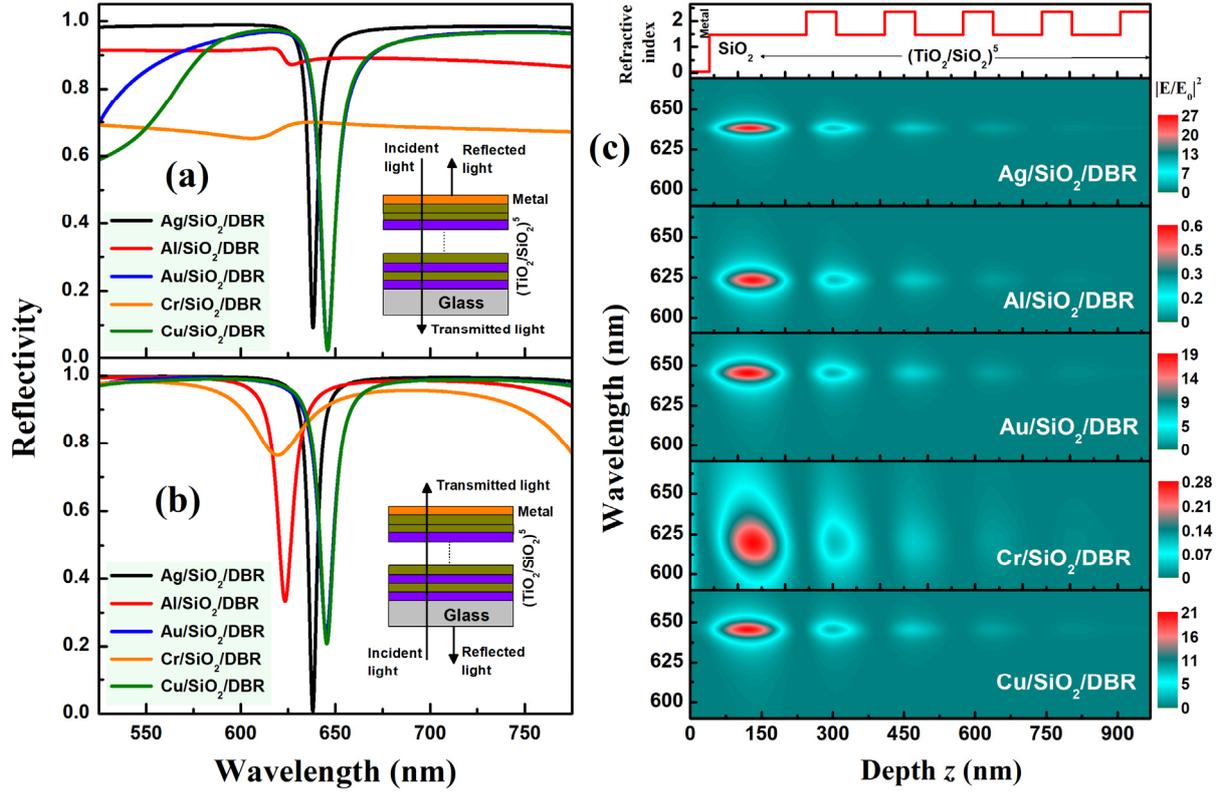

**Fig. 3:** Reflectivity spectra of Tamm plasmon structures (Metal/SiO$_2$/DBR) made of different metals when the light is incident normally from (a) air-metal interface side, and (b) substrate-DBR interface side, respectively. The structure consists of a SiO$_2$ spacer layer placed between metal and a DBR composed of 5 pairs of TiO$_2$ and SiO$_2$, respectively. (c) Wavelength resolved electric field distribution throughout Metal/SiO$_2$/(TiO$_2$/SiO$_2$)$^5$ structures (refractive index profile shown on the top of this plot) made of different metals for the light incident from air-metal interface side.

The behaviour of TP mode for different metals can be well understood by analysing the electric field distribution of the resonant mode in the structure. The wavelength resolved electric field distribution for different metal based TP structures are plotted in Fig. 3(c). It shows that the electric field is localised within the SiO$_2$ layer close to the metal layer. The metallic layer and the DBR act as two reflecting mirrors, which makes the whole structure similar to a Fabry-Perot cavity resulting the resonance at the SiO$_2$ layer. Such amplification of the electric field generates high Q-factor resonant modes. The electric field intensity is much higher and strongly localized for the Ag/SiO$_2$/DBR structure while it is least and weakly localized for the Cr/SiO$_2$/DBR structure. It indicates that the full width half maximum (FWHM) of the TP mode will be narrower and its reflectivity minimum will be deeper for the Ag based structure among all the TP structures. It means the Ag based TP structure will exhibit highest Q-factor which can be clearly seen in the Fig. 3 (a) and (b), respectively. It will be useful for high sensitivity and large detection accuracy sensing applications. The response of TP mode in Metal/SiO$_2$/DBR structure for two most widely used metals Ag and Au are presented in Fig. 4 with varying angle of incidence (AOI) for TE and TM polarised light. With increase in the AOI, the TP mode in both the structures undergoes blue shift for both the polarisations of light. The shift of the TP mode with increasing AOI is more for TE polarised light as compared to that of TM polarised light for both the structures. The Q-factor of the TP mode in Ag/SiO$_2$/DBR structure increases and decreases with increasing angle of incidence (AOI) for TE and TM polarised light, respectively as seen in Fig. 4 (a) and (b). For Au based TP structure, the FWHM of the TP mode increases with increase in AOI for both the polarisations of light as seen in Fig. 4 (c) and (d). From above study it can be concluded that Ag is the best choice for excitation of TPPs among the noble metals. Apart from the noble metals, a number of materials that exhibit very low dielectric constants can also be used to generate TPPs. Recently, TPPs are observed in structure made of materials like transition metal nitrides [66], topological insulators [67], hyperbolic metamaterials [68], graphene [69], black phosphorous [70], and epsilon-near-zero materials [71]. Metal nitrides like TiN has been considered as plasmon active material for generating high Q Tamm plasmon resonances from visible to IR region. Kumar [72] has studied the behaviour of Tamm modes at the interface between TiN and a 1DPC. Yang *et al.* [73] experimentally found that the thermal emissivity and threshold temperature of TiN based TPP structure is higher compared to TiN based metal-insulator-metal (MIM) structure.

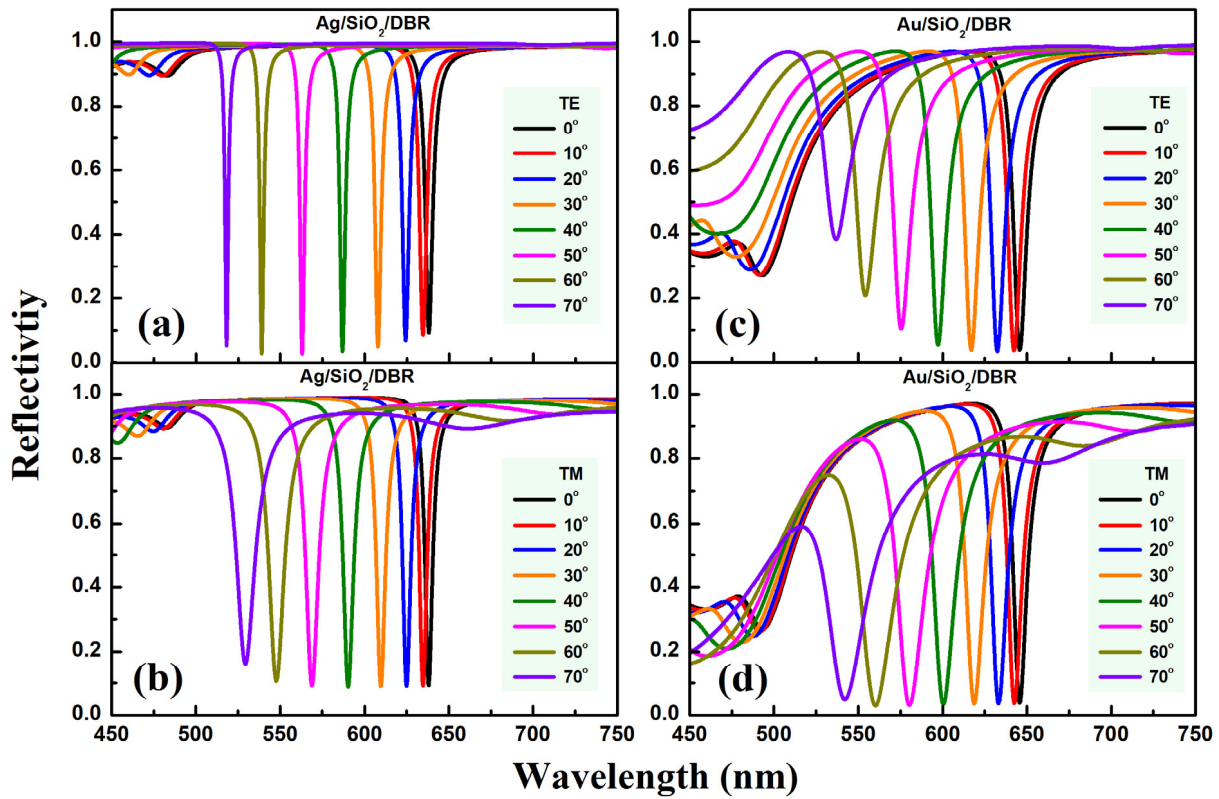

**Fig. 4:** Reflectivity spectra of Ag/SiO$_2$/DBR and Au/SiO$_2$/DBR Tamm plasmon structures for (a, c) TE and (b, d) TM polarised light incident at different angles of incidence.

TPP has unique ability to undergo direct optical excitation by both TE- and TM-polarized light without the aid of prisms or gratings. Here, the effect of metal thickness on excitation of the TPP mode and other resonant modes in a prism coupled TP structure has been illustrated in Fig. 5. A hybrid photonic structure made of a 1DPC and a SiO$_2$ layer with a silver layer placed on its top is shown in Fig. 5 (a). The 1DPC structure is made of six periodic bilayers of (TiO$_2$/SiO$_2$). The thickness values of TiO$_2$ and SiO$_2$ layers are 58 nm and 85 nm, respectively. The dispersion free refractive index values of TiO$_2$ and SiO$_2$ are 2.36 and 1.47, respectively. The structure shown in Fig. 5(a) has been used to investigate the effect of metal (Ag) layer thickness on the TPP mode and its coupling with SPP and/or cavity mode excited by TE and TM polarised light. For the excitation of coupled TPP-SPP mode, a total internal reflection configuration of spectrometer with a 45º prism coupler has been used. The calculated reflectivity contour of the prism coupled hybrid structure as a function of silver thickness and wavelength of light at normal incidence (θ=0º) is plotted in Fig. 5(b). It shows that the TPP mode is very weakly excited for Ag layer thickness below 50 nm and the mode undergoes blue shift with increasing silver thickness up to 50 nm. The TPP mode does not sustain for very thin film of thickness below 15 nm. The mode originates due to the electric field confinement by the metal because of its negative dielectric constant. The effective dielectric permittivity of metal-air system will not be negative enough for very thin metal layer to confine the electric field. But with increasing Ag thickness, the effective dielectric permittivity of the metal-air system will be dominated by the metal film, thus supporting the TPP mode [58]. The Q-factor of the TPP mode and its energy remain constant for Ag thickness more than 50 nm. For high Q-factor TPP mode, the Ag layer of more than 60 nm thick is preferred [59]. The reflectivity of the light corresponding to the TPP mode (wavelength of ~538 nm) has been computed as a function of Ag thickness and angle of incidence of both TE and TM polarised light to reveal the co-existence of TPP and SPP modes simultaneously in the hybrid structure. The reflectivity of the structure as a function of Ag thickness and angle of incidence of an excitation wavelength of 538 nm is plotted in Fig. 5(c). It can be seen that there exist SPP mode at the Ag-air interface for a certain Ag thickness range of 20-70 nm for the structure excited by TM polarised light. The resonant angle of the SPP mode and its band width decrease with increasing Ag thickness. The strong resonance angle of the SPP mode is around 44.5º which is larger than the total internal reflection angle of the glass-air interface (41.2º). As the Ag thickness increases the SPP mode becomes less and less prominent due to high propagation loss and finally disappears. Cavity mode with resonant angle of around 37.5º is noticed for the structure excited by TM polarised light which implies that there is transmission of light through the structure at these incidence angles. The band width of the cavity mode is larger than those of SPP. The cavity mode disappears with the increase of the Ag film thickness due to the large reflectivity of the thick Ag film. No such SPP or cavity modes are observed for TE polarised light. It confirms that TPP mode can be excited by both polarised states while the SPP mode can only be excited by the TM polarised light. Fig. 5 clearly demonstrates the relation between Ag film thickness and the confined optical modes existing in the hybrid structure.

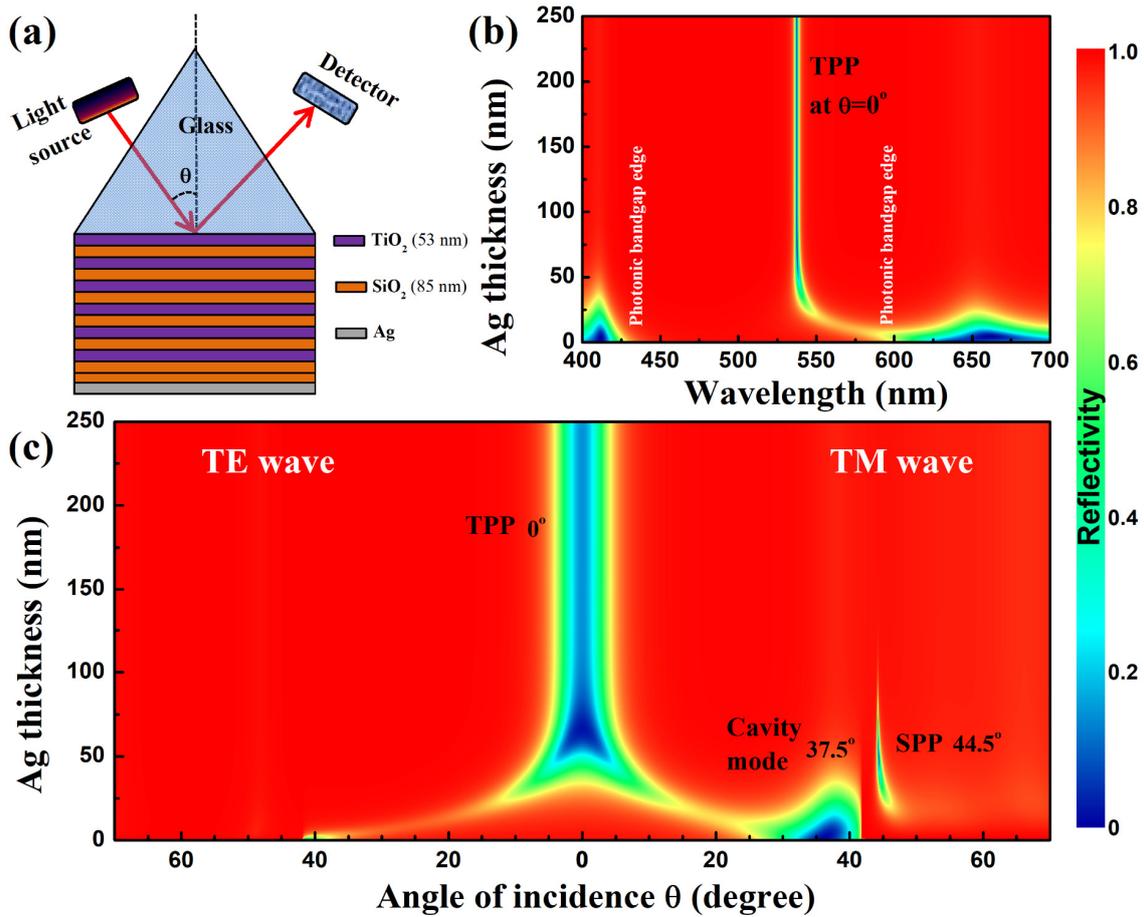

**Fig. 5:** (a) Schematic of prism coupled hybrid photonic structure that supports TPP and SPP modes. (b) Wavelength and Ag thickness resolved reflectance contour. (c) Reflectivity contour as a function of Ag thickness and angle of incidence of light having an excitation wavelength of 538 nm.

## 5. Coupling of Tamm plasmon with other resonant modes

Coupling of Tamm plasmon polariton with other resonant modes can be utilized to tune the resonances. The TPP mode has been coupled with surface plasmon polaritons or magnetic plasmon polaritons for enhancing electric field or magnetic field [69], cavity modes for narrowband thermal emission [74], semiconductor exciton in polaritonic devices [75], and defect modes for induced transparency [76]. The coupling of two or more resonant modes generate "hybrid" modes, which are observed as a series of non-overlapping dips or peaks in the reflection or transmission spectra as shown in Fig. 6. The hybrid modes show Rabi-like splitting in the strong coupling regime. The splitting modes can be used for developing self-referenced refractive index sensors. Fig. 6 illustrates an overall demonstration of the coupling of Tamm plasmon modes to various resonant modes along with their regime of strong coupling. Schematic of the multilayer hetero-structure resulting coupled TPP-cavity modes is shown in Fig. 6(a). The multilayer configuration for the structure is Ag/S/(HfO$_2$/SiO$_2$)$^5$/C/(HfO$_2$/SiO$_2$)$^5$. Both HfO$_2$ and SiO$_2$ layers are assumed as lossless and dispersion free dielectrics with refractive index values 1.96 and 1.47, respectively. Both spacer (S) and cavity (C) layers are made of $SiO_2$. Here, the spacer layer thickness ($d_s$) can be adjusted to control the coupling between the TPP and the cavity modes as shown in Fig. 6(b). The figure has curved dotted and solid lines in the contour plot, which represent the bare cavity mode ($\omega_{cavity}$), bare TP mode ($\omega_{TP}$), and their coupled modes ($\omega_U$ and $\omega_L$) as a function of $d_s$, respectively. The bare TP mode undergoes redshift with increasing $d_s$, while the bare cavity mode has a fixed energy value of ~2.486 eV and it does not change with $d_s$. The coupled TP-cavity hybrid modes ($\omega_U$ and $\omega_L$) strongly deviate from their bare modes in the strong coupling region (130 nm ≤ $d_s$ ≤ 150 nm). Beyond this region, the behaviour of the hybrid modes with $d_s$ is similar to that of the bare modes. The anticrossing of the modes is observed for $d_s$=141 nm at the energy of ~ 2.48 eV, which is their common bare resonant mode energy. It indicates the strong coupling of the modes. The energy separation between the two hybrid modes $\omega_U$ and $\omega_L$ is 125 meV exactly at the anticrossing region, known as Rabi-like splitting energy ($\Omega_R$). The strong coupling between the TPP and cavity modes is further confirmed by the splitting energy value as it is larger than the linewidth of their bare modes [77]. Coupled oscillator model can be used to better explain the coupling between resonant modes. Here, the cavity mode and the TPP mode are considered as two oscillators coupled in a hybrid structure. The energy eigenvalues ($\omega$) of the coupled oscillators system has been obtained by solving the a 2x2 matrix as described elsewhere [78]. The two solutions can be expressed as follows:

$$\omega_U(d_s) = \frac{1}{2}\left[(\omega_{cavity} + \omega_{TP}(d_s)) + \sqrt{(\omega_{cavity} - \omega_{TP}(d_s))^2 + 4\Omega_{cT}^2}\right] \quad (10)$$

$$\omega_L(d_s) = \frac{1}{2}\left[(\omega_{cavity} + \omega_{TP}(d_s)) - \sqrt{(\omega_{cavity} - \omega_{TP}(d_s))^2 + 4\Omega_{cT}^2}\right] \quad (11)$$

where $\Omega_{cT}$ is the coupling strength between the cavity and TP modes. The TP mode is tuned by varying the spacer layer thickness $d_s$ close to the Ag layer. It means that the value of $\omega_{TP}$ is a function of $d_s$ i.e. $\omega_{TP}(d_s)$. Equation (10) and (11) have been used to fit the curves of hybrid modes $\omega_U$ and $\omega_L$ as shown in Fig. 6 (b). The fitted curves are exactly matching to the numerically estimated hybrid resonance positions in the reflectivity contour. It validates the effectiveness of coupled oscillator model to explain the behaviour of coupled modes. Fig. 6 (c) shows the reflectivity spectra for different values of $d_s$ which helps to understand the coupling of modes. It can be clearly seen that the hybrid modes are coupled with unequal weight of the TP mode and the cavity mode for $d_s$=180 nm, 160 nm, 120 nm, and 100 nm, which is evident from their different linewidth and reflectivity dip minima in the reflectivity curves. It indicates that the modes are weakly coupled. In case of $d_s$=140 nm, the coupled modes are formed with equal weight of both their bare modes indicating strongly coupled system, which is evident from their identical amplitude and linewidth of the resonant dips in the reflectivity spectrum. It infers that a coupled system can be designed as weak or strong by tuning the $d_s$ value. Here, the spacer layer acts as a modulator for tuning hybrid resonances in the coupled system.

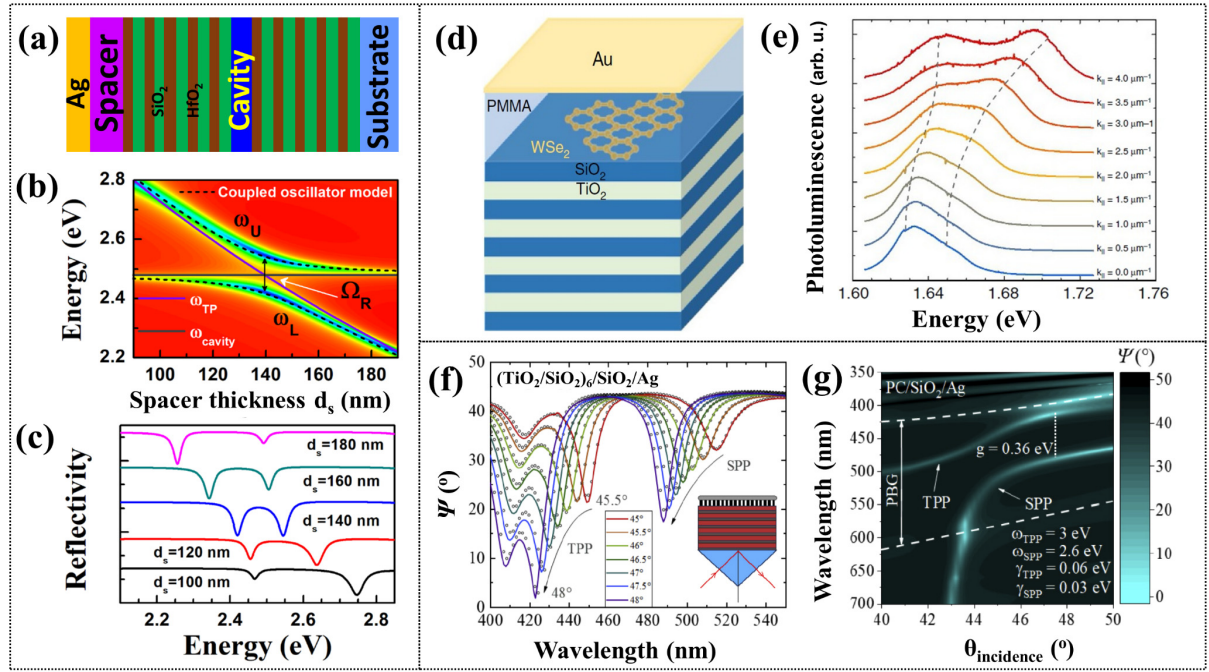

**Fig. 6:** (a) Schematic of a photonic hetero-structure Ag/spacer/1DPC/cavity/1DPC giving coupled TPP-cavity modes, (b) contour plot of reflectivity of the structure as a function of spacer layer thickness ($d_s$) and energy, and (c) reflectivity curves for different values of $d_s$ [54]. (d) Schematic of TPP-exciton coupled system with an embedded WSe$_2$ monolayer and (e) its inverted photoluminescence spectra for various in-plane momenta showing splitting of coupled Tamm plasmon-exciton polariton modes [75]. (f) Ellipsometric parameter $\Psi$ spectra of TPP-SPP hybrid mode excited in a 1DPC/SiO$_2$/Ag structure shown in the inset for different angle of incidence ($\theta_{incidence}$), and (g) contour of $\Psi$ as a function of wavelength and $\theta_{incidence}$, where the white dashed line corresponds to the photonic band gap edges [79]. All the figures are adapted with permission.

The coupled hybrid modes provide a feasible platform for many integrated photonic applications like biosensors, emitters, filters, absorbers, and photodetectors [80-82]. The coupled TPP-cavity modes can be tuned very easily by varying the geometrical parameters and polarisation of the light [54]. Fig. 6(d) illustrates a compact photonic device exhibiting room temperature Tamm plasmon coupling to a valley exciton of WSe$_2$ monolayer embedded in it [75]. The exciton has resonance energy of 1.650 eV with a line width of 37.5 meV. The WSe$_2$ monolayer is placed close to the metallic layer so that the Tamm plasmons formed can effectively couple to the excitonic mode. Experimentally observed photoluminescence spectrum is shown in Fig. 6(e) at various in-plane momenta. The modes evince strong coupling between them with a Rabi splitting of 23.5 meV. Similarly, TPP-SPP hybrid photonic states are formed when the TPP and SPP are simultaneously excited in a photonic hetero-structure as shown in Fig. 6(f) which requires TM polarised light for SPP excitation [83]. The mode hybridization results in energy exchange between the interacting modes so that the dispersion of the new resonant modes is modified compared to the non-interacting isolated states. This can be observed in the momentum or angle resolved reflection-transmission spectra as shown in Fig. 6 (g), It shows a Rabi-like splitting between the hybrid modes in the strong coupling regime [84]. The coupling strength between SPP and TPP modes can be estimated using coupled oscillator model [79] and is given by

$$g = \sqrt{\Omega_R^2 - \frac{1}{4}(\omega_{SPP} - \omega_{TPP})^2} \qquad (12)$$

In the strong coupling regime $g > (\Upsilon_{SPP} + \Upsilon_{TPP})/4$, where g is the coupling strength, $\Omega_R$ is the Rabi splitting between the modes, $\omega_{TPP/SPP}$ is the resonance frequency of TPP/SPP mode, $\Upsilon_{TPP/SPP}$ is line width of TPP/SPP mode.

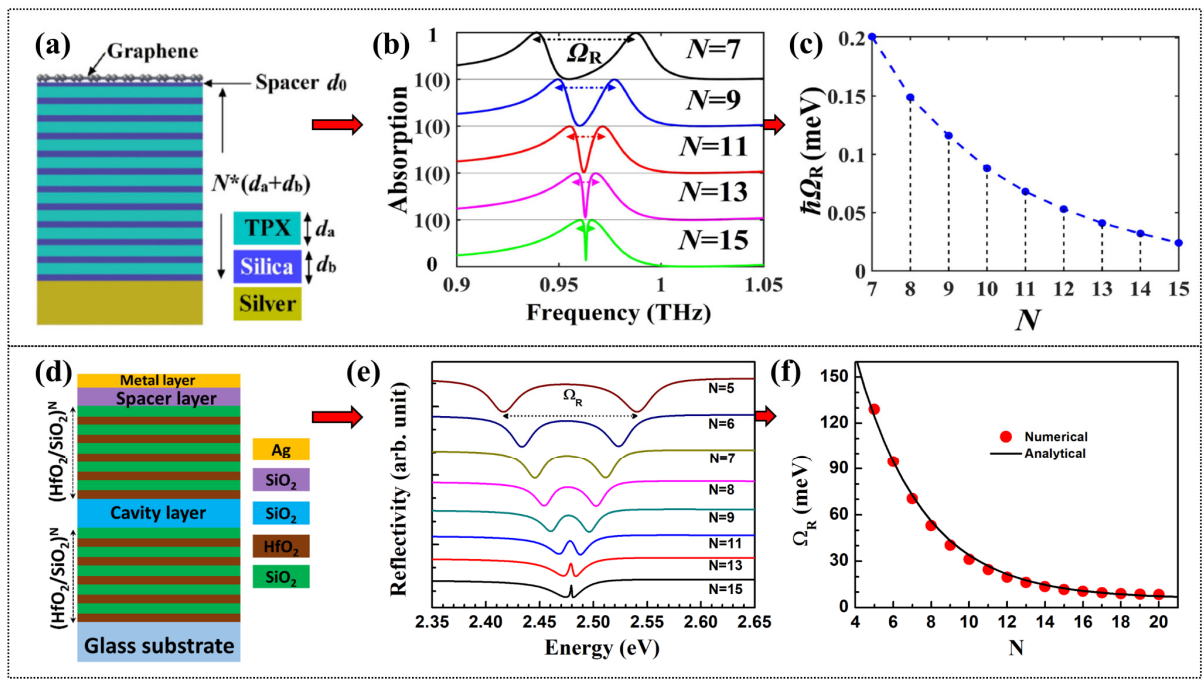

**Fig. 7:** (a) Schematic of a graphene based coupled modes structure, and its (b) absorption spectra and (c) Rabi-like splitting energy ($\Omega_R$) with varying periodicity (N) of the DBR [69]. (d) Geometric configuration of a hybrid TPP-cavity modes structure, and its (e) reflectivity spectra and (f) $\Omega_R$ with varying N [54]. All the figures are adapted with permission.

It is found that the coupling of hybrid modes depends upon the periodicity of a DBR in a TPP configuration. Fig. 7(a) depicts a device configuration exciting graphene Tamm plasmon polariton at the graphene-DBR interface and silver Tamm plasmon polariton at the silver-DBR interface [69]. The change in coupling of these two modes with varying DBR periodicity can be clearly seen from Fig. 7(b) and (c). When the graphene layer placed next to metallic layer, it changes the conductivity of the metal layer which modifies coupling strength between the modes [79]. Another schematic of a device exhibiting coupled TPP-microcavity modes [54] is shown in Fig. 7(d) whose coupled mode behaviour with varying DBR periodicity is also presented in Fig.7(e) and (f). The study shows that the behaviour of modes coupling with DBR periodicity will remain same even though the device configurations are different *i.e.* the Rabi-like splitting energy between the modes decreases with increasing periodicity and the modes gets narrower which can be attributed to the reduced radiation loss through the DBR. The value of $\Omega_R$ in the strong coupling region can be calculated using the following equation [54]:

$$\Omega_R = 2\Omega_{cT} = \frac{2}{\pi}\left(\frac{1}{\eta}\right)^N \left[\frac{1-(1/\eta)}{\sqrt{1-(1/2\eta)}}\right]\omega_0 \tag{13}$$

where $\eta = n_{HfO2}/n_{SiO2}$ is the refractive index contrast, and $N$ is the periodicity or number of unit cells in a 1DPC. From equation (13), one can clearly see that the value of $\Omega_R$ does not depend on the properties of the metal layer. Equation (13) has been used to compute the values of $\Omega_R$ with varying $N$ and it is plotted in Fig. 7 (f). The computed curve is closely matching with that of the values estimated numerically using transfer matrix method. The value of Rabi-like splitting energy $\Omega_R$ is large (~125 meV) for the lower value of $N=5$. The resonant electric field of the TP mode easily propagates through the 1DPC made of lower $N$ value, and efficiently interacts with the cavity mode resulting in a large value of $\Omega_R$. The strong coupling between the cavity mode and the TP mode leads to an exchange of energy between the cavity layer and metallic silver interface. Consequently, enhanced light-matter interaction occurs in either of the cavity and spacer layers.

Recently, coupled TPP-guided mode resonance [85], Tamm state-plasmonic defect mode [86], TPP- magnetic plasmon [87], and TPP-topological photonic state [77] have been studied to observe the generated multiple hybrid photonic states useful for different photonic applications. Tamm coupled fluorescence emission offers an interesting area of research [88-90]. In a study, directional control of fluorescence emission by use of single nano-aperture on a metallic film deposited on a DBR has been experimentally observed. The fluorescence emission coupled to Tamm plasmon at metal-DBR interface is observed to be emitted with an angular width of 12.4° from the sample surface [90]. Coupled Tamm plasmon-exciton polaritonic states are often exploited for realisation of enhanced spontaneous emission and compact polariton lasers [91]. The primary reason for enhancement of spontaneous emission is the high confinement of field near to the interface leading to better light matter interaction. Not only coupled TP structures assist to enhance spontaneous emission but also it is easy to control the emission pattern by modifying the metallic layer of the planar metal-PC structures [92]. Advance studies include the coupling of Tamm plasmons with 2D excitons for compact photonic devices [93]. Strong coupling between TP and exciton modes of QWs was found to produce intense emission of polaritonic modes useful for polariton lasers [36, 39]. It is reported that TPs can be used mediate to couple quantum dot emission to SPs due to its deeper penetration into the DBR [94].

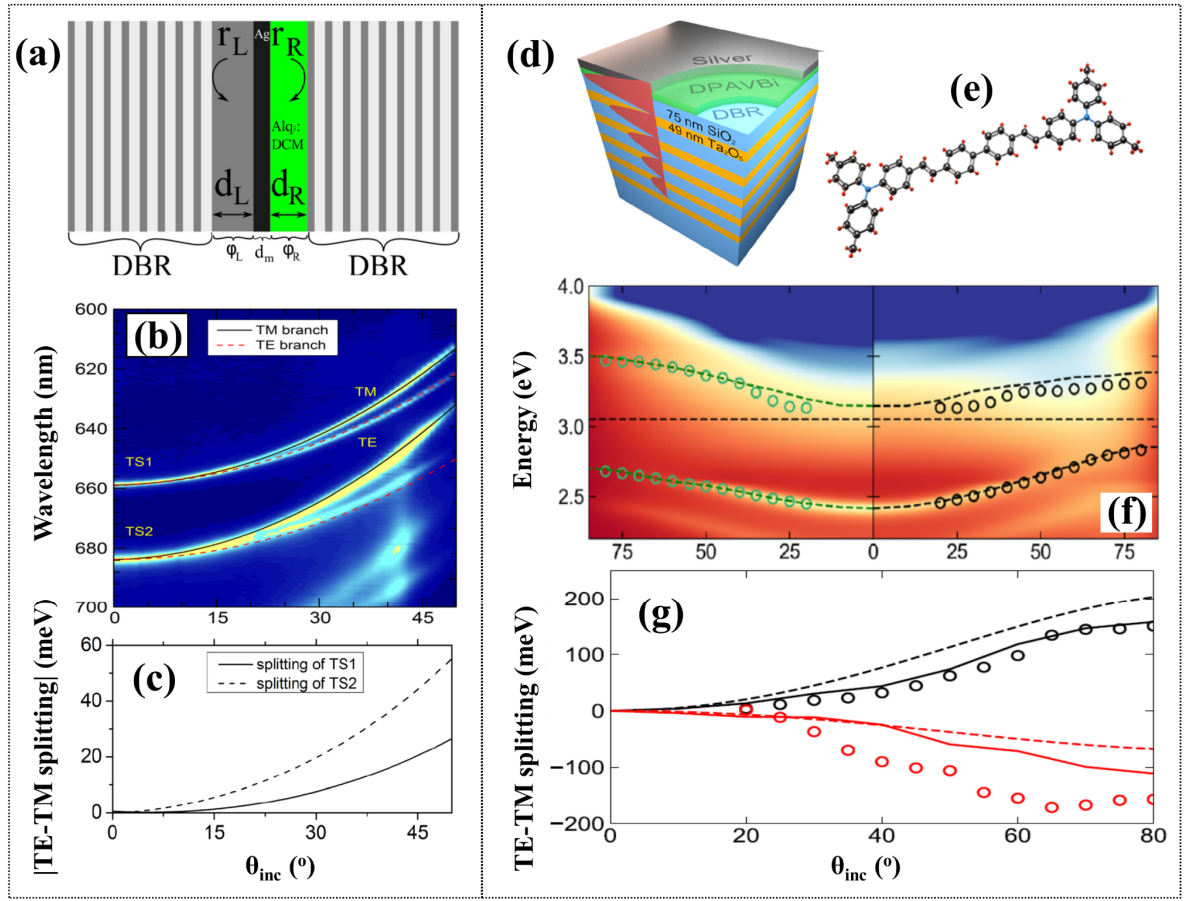

**Fig. 8:** (a) Sketch of a structure containing two Bragg reflectors enclosing an organic cavity layer and a silver layer that generates hybrid modes of organic microcavity photon and Tamm plasmon polariton, (b) measured angle resolved emission spectra of the hybrid modes along with the theoretical data (black solid and red solid line) for TE and TM polarized light, and (c) parabolic splitting between TE- and TM-modes of the shifted cavity resonance and Tamm plasmon-polariton with varying angle of incidence [95]. (d) Schematics of a hybrid mode structure containing strong oscillator layer made of (e) DPAVBi molecules, (f) calculated (dashed line) and measured (circles) reflectivity spectra of the structure having 60 nm thick DPAVBi layer for TE (left) and TM (right) polarized light, and (g) experimentally (circle), analytically (dashed line), and numerically (solid line) obtained polarization splitting of the hybrid modes [96]. All the figures are adapted with permission.

## 6. Polarization splitting of Tamm plasmon coupled resonant modes

Angle and polarization resolved reflectivity spectra of a coupled resonant modes provide detail information on interactions between modes [97]. Polarization splitting, which is the energy difference between TE and TM polarised light excited resonant mode, can be determined from the measured angular reflection/emission spectrum. It is also called as TE-TM splitting. In optical micro-cavities, the TE-TM splitting enables the observation of interesting phenomena including the optical spin Hall effect, magnetic monopole-like half solitons, spinor condensate with half-quantum circulation, and possibly topological insulators [98]. In case of optical spin hall effect, the spin polarization of the polaritons scattered clockwise and anticlockwise have different signs, which is possible due to the strong TE-TM splitting and finite life time of the exciton polariton in micro-cavities [99]. Strong coupling between TPPs and other excitons leads to coupled resonant modes and their polarization dependency can be explored for optoelectronic device applications [95]. A photonic nanostructure device has been shown in Fig. 8(a) comprising an organic cavity and a Ag layer sandwiched between two DBRs [95]. The structure leads to coupled Tamm plasmon states (TS1 and TS2). The angle resolved reflection spectra of the coupled modes for both TE and TM polarised light are shown in Fig. 8(b). The resonant coupled modes get blueshift with increasing incident angle of light. It happens due to the resonance condition $\delta=2\pi nd\cos\theta/\lambda$, where $d$ is the virtual cavity thickness. As per this equation, the resonant mode energy has to increase to keep a fixed phase shift ($\delta$) with increasing incident angle ($\theta$). The linewidth and resonant dip of both the modes are constant over a wider range of incident angle, which makes it suitable for developing tunable dual-narrow-band filters. The coupled modes exhibit parabolic polarisation splitting of more than 40 meV confirmed from the analytical dispersion curves. Fig. 8(c) verifies the quadratic splitting of the distinct modes and the angular behaviour of the splitting in the strong coupling regime can be well described by the $sin^2\theta$ dependency predicted by the polarization splitting model [100]. The lower energy Tamm plasmon state TS2 exhibits large TE-TM splitting as compared to that of the higher energy state TS1, which can be attributed to the plasmonic nature of the TM mode [101]. There is also a report indicating opposite sign of polarisation splitting in organic microcavity-Tamm structures in the regime of ultra-strong coupling [96]. The structure shown in Fig. 8(d) is used for ultra-strong coupling of Tamm plasmons and exciton of DPAVBi (4,4′-Bis[4-(di-ptolylamino) styryl]biphenyl) whose chemical structure is shown in Fig. 8(e). The analytical, numerical and experimental data of angle resolved emission spectrum are shown in Fig. 8(f) for a DPAVBi layer thickness of 60 nm. The value of TE-TM polarisation splitting is more than 180 meV for both upper (UP) and lower polariton (LP) branches as shown in Fig. 8(g), which is much larger to the earlier reported results. Panzarini *et al*. [102] have reported the maximum TE-TM splitting of ~1.7 meV for InGaAs quantum wells in a GaAs cavity. Lodden and Homes [103] have demonstrated a significant polarization splitting of less than 30 meV in an organic semiconductor microcavity under both optical and electrical exciton. Camposeo *et al*. [104] have reported the maximum TE-TM splitting of 35 meV for *J*-aggregate of cyanine dyes as active layer in an organic microcavity. Hayashi *et al.*

[101] have shown a maximum TE-TM splitting of ~150 meV in a metal-insulator-metal microcavities with a PVA layer doped with *J*-aggregates of TDBC molecules as an active layer. Oda *et al.* [105] have reported the maximum polarization splitting of ~200 meV for strong exciton-photon coupling in a metal mirror microcavity with oriented PIC *J*-aggregates as an active layer. The reported data suggests that the polarization splitting can be exploited for manipulating, shaping and guiding propagation and confinement of coupled Tamm plasmon resonant modes for realizing Tamm plasmon based optoelectronic devices [106].

## 7. Tunable Tamm plasmon

Tamm plasmon mode excited at the interface between a metal and a DBR does not get affected by any perturbation in the external environment, but it can be tuned by changing properties of active layers. Tunable TP modes has plenty of applications including spatial filters, broad range optical sensors, and solar power harvesting [107]. Most common approach to get tunable TP mode is by changing the thickness of layer and number of periods in DBR, varying thickness of metal layer, etching microstructure on the metal films, inserting 2D materials in the DBR, and gradient thick metal layer [35, 108-110]. Active tunability of TP mode by applying external electric field, magnetic field, temperature, and pressure have been demonstrated by several researchers in recent years. Chen *et al.* [110] have tuned resonant wavelength, band width, and Q-factor of the coupled Tamm plasmon-cavity modes by varying spacer thickness, fermi energy of graphene, Bragg grating period, and Bragg wavelength. The surface conductivity of the graphene layer depends on its fermi energy which can be controlled by applying electric field or heat [111]. It is exploited for tunable Tamm plasmon mode. Li *et al.* [112] have theoretically demonstrated tunable reflectivity dip modulation from 38% to 99% by applying an electric field in a TPP structure with a monolayer graphene. Mode hybridization of Tamm plasmons has been preferred to design actively tunable TP devices. Such tunability can be achieved by inserting a liquid crystal (LC) layer in a conventional TP structure [113, 114]. Buchnev *et al.* [46] have dynamically tuned a Tamm structure by applying voltage to a LC above a metasurface. The inclusion of the LC turns the hybrid modes sensitive to heat, electric field and polarization of the incident light. The mode depends upon the applied electric field because of the Frederiks transition in the LC [115]. Fig. 9(a) shows a Tamm plasmon device in which a LC 4-pentyl-40-cyanobiphenyl (5CB) is inserted between two PCs whose cross-sectional scanning electron microscope (SEM) image is shown in Fig. 9(b) [113]. The modes changes significantly for TM polarization with the applied voltage after the Frederik threshold voltage of 0.74V which can be seen in the measured reflectance spectrum plotted in Fig. 9(c).

No change occurs for the TE polarized light because the external electric field is applied along the direction of propagation. The molecular ordering of the LC molecules decreases with increase in temperature which leads to decrease in extraordinary refractive index and increase in ordinary refractive index of the LC. As a result, the micro-cavity mode undergoes blue shift and red shift, respectively for TM and TE polarized light. Measured and calculated reflectance spectra with temperature for TM polarization is plotted in Fig. 9(d) showing the anti-crossing between the TPP-MC hybrid modes. The shift in the hybrid modes happens due to the change in refractive index of the liquid crystal after nematic-isotropic phase transition at 33.4ºC. Another LC assisted actively tunable TP device is shown in Fig. 9(e) [114]. The reflectance spectra of the device plotted in Fig. 9(f), (g) and (h) reveal the tunability of Tamm mode by varying the angle of incidence to 0º, 45º, and 90º, respectively due to the birefringence property of the LC layer. LC based tunable Tamm plasmon devices are widely demonstrated due to their large birefringence and easy controllability via external stimuli [116-118].

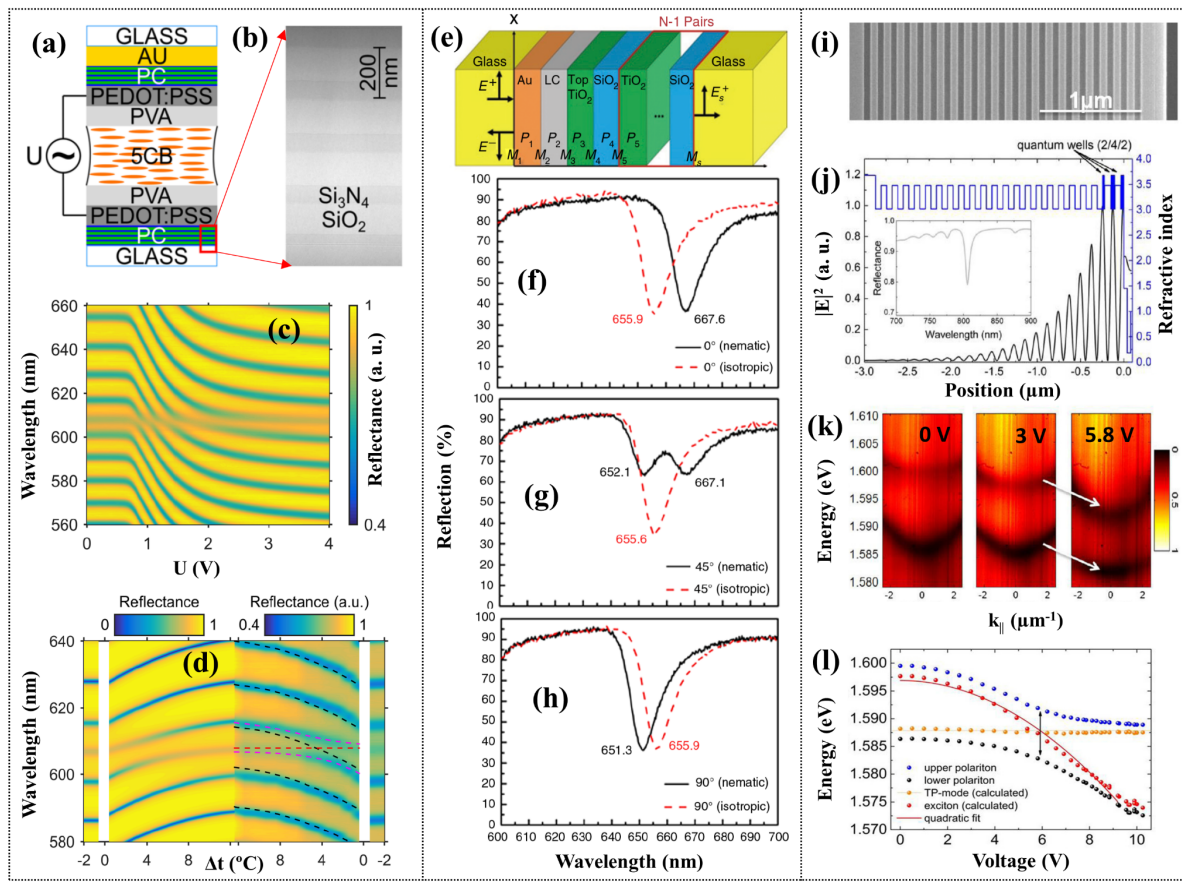

**Fig. 9:** (a) Schematic of the liquid crystal (LC) micro-cavity Tamm structure, (b) electron microscope image of the 1DPC made of periodic layers of $SiO_2$ and $Si_3N_4$, (c) reflection contour as a function of externally applied electric potential and wavelength of light at 21.8ºC, and (d) calculated (left) and measured (right) temperature dependent reflectance spectra for TM-polarized light incident at an angle of $4^0$. The red dashed line and black dashed line correspond to the position of the bare TPP and bare micro-cavity modes, respectively whereas the magenta dashed line corresponds to the results from coupled oscillator model [113]. (e) Sketch of TPP structure made of a LC layer sandwiched between a 1DPC made of $TiO_2/SiO_2$ periodic layers and Au coated glass substrate,

and its reflectance spectra measured at incident angle of (f) 0º, (g) 45º, and (h) 90º, respectively for the LC layer switching between the nematic and isotropic phases [114]. (i) SEM image of a structure consisting of a bottom($Al_{0.2}Ga_{0.8}As$/AlAs)$^{20}$DBR mirror and atop layer of $Al_{0.2}Ga_{0.8}As$ embedded with eight GaAs quantum wells, (j) electric field intensity and refractive index profile inside the structure with its reflectivity spectrum shown in the inset plot, (k) momentum resolved reflectivity spectra of the structure at different indicated voltages, and (l) variation of hybrid modes (lower and upper polariton) with applied voltage exhibiting Rabi-like splitting are plotted along with the bare TP mode and the bare exciton mode [119]. All the figures are adapted with permission.

Electro-optical tuning is another convenient and reversible methods used for tuning of the modes. Grossmann *et al.* [36] have demonstrated electro-optical tuning of Tamm modes in an air-gap DBR micro-cavity structure possessing quantum well excitons. The strong coupling between exciton-Tamm plasmon modes is actively tuned by electrical biasing. In another study containing gold micro disks, the hybrid exciton-polariton modes show red-shift under external electrical field [119] whose cross-sectional SEM image is shown in Fig. 9(i) and corresponding field profile and reflectance spectrum is given in Fig. 9(j). The fields are highly amplified in the QW reasons leading to strong coupling between the modes. The voltage dependent dispersion is calculated from the momentum resolved reflectance shown in Fig. 9(k). With the increase in voltage, the emission shifts red while the dispersion curvature of both the upper and lower branch flattens induced by the quantum confined Stark effect. Under electrical biasing a Rabi splitting of (9.2 ± 0.2) meV is observed between the modes plotted in Fig. 9(l). The Rabi energy is useful for calculation of oscillator strength per unit area of the QW. It is evident that the hybrid system can change from highly excitonic to highly polaritonic state under electrical tuning as observed in Fig. 9(l). Bikbaev *et al.* [120] have proposed a Tamm structure containing metagrating and a conductive oxide ITO to dynamically control the amplitude and phase of the reflected light under electrical biasing. The TP resonance undergoes blue shift with increase in applied voltage due to decreasing dielectric permittivity of the ITO layer. Dong *et al.* [121] have theoretically demonstrated non-reciprocal TPPs at the interface of a magneto-photonic crystal (MPC) and a conducting metallic oxide. They observed spectral splitting of the TPP modes by front and back illumination which validates the non-reciprocity arising due to violation of periodicity and time-reversal symmetry in the structure. He *et al.* [122] have reported nonreciprocal resonant transmission/reflection, which originates from direct excitation of nonreciprocal TPPs at the interface between a 1DPC and a magneto-optical metal film by applying an external magnetic field. The magneto-optical thin film based TP structure will be useful for designing optical nonreciprocal devices such as optical diodes. Wu *et al.* [123] have observed strong nonreciprocal radiation at $30^0$ angle of incidence in a MPC atop silver layer in which the MPC consists of periodic layers of InAs and a spacer layer with relative permittivity of 2. TPPs possess many interesting characteristics in the magnetic domain which require more theoretical and experimental investigation. Buller *et al.* [124] have studied the formation of Tamm plasmon/exciton-polariton hybrid states and their modulation by applying surface acoustic waves in a structure made of a gold layer on top of a DBR consisting of $Al_xGa_{1-x}As$ layers. The modulation occurs due to the change of the exciton band gap energy and change of layer thickness because of induced strain fields by the surface acoustic waves. Active modulation of Tamm plasmon modes is a very hot topic and requires numerous efforts for cost effective, reversible and durable Tamm devices for their application in various fields.

## 8. Applications of Tamm plasmon

Tamm plasmons are being considered for innovative applications in perfect absorbers, nanoscale lasers, filters, bistable switches, sensors, thermal emitters, solar photovoltaic cells, photodetectors, photocatalysis, and many other applications. This section presents brief summary of the progress made in some of the applications.

### *8. 1. Perfect absorbers*

There has been increasing interest in making photonic nanostructures for perfect absorption at designated wavelengths with potential applications in solar cells, lasers, integrated photonics etc. Gong *et al.* [125] designed a near perfect absorber exploiting Tamm modes in a 1D photonic crystal made of alternate layers of $TiO_2$ and $SiO_2$ in a 2D metal–dielectric-metal waveguide. Thin metallic layer near the PC excited the Tamm modes to exhibit near perfect absorption (99.1%) whose peak is tunable from telecom wavelengths (1550 nm) to visible (590 nm) by changing the geometrical parameters of the structure. Liu *et al.* [126] have theoretically and experimentally demonstrated single and multi-band near perfect absorber by using simple metal and a 1D PC based structure. Perfect absorbers for visible region are observed in a simple photonic heterostructures composed of a truncated all-dielectric photonic crystal and thick metal film [127] whose schematic is shown in Fig. 10(a). For absorption purpose, small thickness of the metal layer is detrimental. The numerical simulation reveals that absorption more than 99% is possible for silver thickness of more than 150 nm with an eight period DBR structure. The influence of different period numbers (N) on the absorption of the structure are investigated. The measured absorption spectra of the structures and their SEM images are shown in Fig. 10 (b), (c) and (d) for N=7, 8 and 10, respectively. The three structures exhibit near perfect absorption of 92.3%, 90.1%, and 90.6% at wavelengths of 675 nm, 604 nm, and 489 nm, respectively. Lu *et al.* [128] demonstrated high absorption of 80% at telecom wavelengths for both TE and TM light excitations on introduction of a monolayer graphene between a DBR and a thin metal layer. In a similar design, a monolayer MoS$_2$ has been introduced between a dielectric Bragg grating and a metal film. The resulting Tamm mode in the photonic structure enhanced the light absorption by 96% in the visible region [129]. Han *et al.* [130] demonstrated graphene based tunable mid IR absorber structure exploiting coupling of Tamm plasmon polariton and surface plasmon polariton. Their structure comprises of DBR, air layer, SiC, and Graphene ribbons. Interaction of Graphene localized SPP and TPP modes lead to dual mode IR absorption above 99%. Confinement of Tamm modes lead to high absorption in 2D graphene layers which otherwise absorb only 2.3% incident light. There are many reports on graphene assisted Tamm plasmon based tunable absorbers form visible to infrared region of the electromagnetic spectrum [85, 131-133].

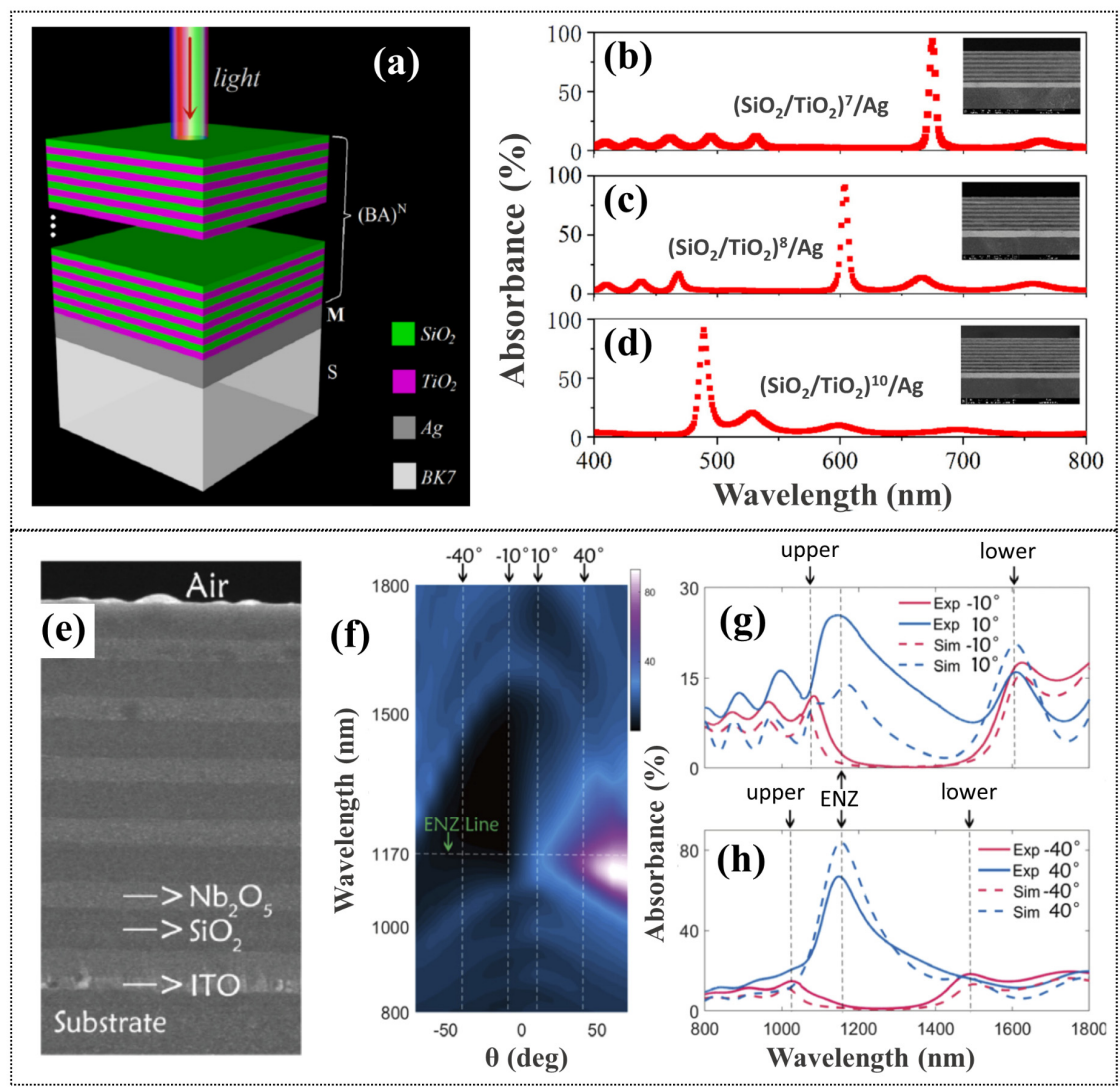

**Fig. 10:** (a) Schematic (TiO$_2$/SiO$_2$)$^N$/Ag Tamm plasmon structure as a perfect absorber, and measured absorption spectra of the (TiO$_2$/SiO$_2$)$^N$/Ag structures with their cross-sectional SEM images (inset plot) for (b) N=7, (c) N=8, and (d) N=10, respectively [127]. (e) Cross-sectional SEM image of ITO/(Nb$_2$O$_5$/SiO$_2$)$^6$ heterostructure based non-reciprocal Tamm absorber, (f) measured absorption map in the heterostructure as function of wavelength and angle of incidence for TM polarised light, and comparison plots between simulated and measured absorption spectra at (g) ±10$^0$ and (h) ±40$^0$ angle of incidence [71]. All the figures are adapted with permission.

SEM image of an angle insensitive Tamm absorber using epsilon near zero (ENZ) material [71] is shown in Fig. 10 (e). Doped indium-tin-oxide (ITO) of thickness 60 nm as ENZ material is used to replace metal as shown in the SEM image. The 1DPC is composed of $Nb_2O_5$ and $SiO_2$ of thicknesses 150 nm and 215 nm, respectively. The real part of the permittivity of ITO is close to zero ($\varepsilon_{ITO}'=0.0037$) at $\lambda \approx 1170$ nm. It is a signature of an ENZ material, and leads to absorption peak at $\lambda \approx 1170$ nm for TM polarised light in ITO thin film. The absorption of the Tamm structure as a function of wavelength and angle of incidence for TM polarised light is shown in Fig. 10(f). The positive angle, in this case, corresponds to the clockwise angle of incidence when light is incident from the substrate side and negative angle corresponds to the anticlockwise obtained angle of incidence when the light is incident from the air side. The non-reciprocity behaviour of absorption is clearly visible in Fig. 10 (g) and (h) when TM polarised light is incident at $\pm 10^0$ and $\pm 40^0$ respectively. Large nonreciprocal absorption around the ENZ position *i.e.* $\lambda \approx 1170$ nm is observed. The intensity of the absorption for different angle of incidence is found slightly different, but the nonreciprocal nature always exists because of the ENZ effect. Substantial absorption can be observed in Fig. 10 (h) when the light is incident from the substrate side or ITO side at $40^0$ which can be explained as the cumulative effect of confinement of the Tamm plasmon and the boundary condition induced localization at the ENZ interface. Xue *et al.* [134] designed a wide-angle spectrally selective absorber with 98% absorption over a wide range of incidence angles up to 80º, where the dispersion-less TPP mode is excited in a structure made of 1D hyperbolic metamaterials on metal substrate. An angle insensitive broadband absorber is designed by Wu *et al.* [135] with absorbance of more than 85% from 1612 to 2335 nm. The device is based on Tamm plasmon excitation between Cr metal and a 1DPC containing hyperbolic metamaterials with a wide angle absorption from 0 to 70º. Lu *et al.* [136] have simulated and experimentally demonstrated a wide-angle perfect absorber in a heterostructures made of a metal layer and a 1DPC composed of layered hyperbolic metamaterials and dielectrics. The absorption peak with TPP shows value of 91% in the angle range of 0 to $45^0$. Li *et al.* [109] has designed a metal-photonic crystal structure that is tunable over telecom range (1510–1690 nm) exploiting Tamm plasmons. Their optimized structure has shown absorbance of 99.99%. Bikbaev *et al.* [137] have theoretically demonstrated a narrowband perfect absorber exploiting TPP localized at the interface between a 1DPC and a nanocomposite with near-zero effective permittivity. Recently, Kim *et al.* [138] have proposed a TP structure made of single material whose refractive index is tailored through porosity and prepared using glancing angle deposition technique. The single material TP structure shows near-unity absorption of ~99% with Q-factor of ~45 at $\lambda$=700 nm. It is evident that TPPs have been extensively studied to realize wide angle, tunable, narrowband/broadband, and multichannel absorbers useful for various nan-photonic devices [139-141].

## 8. 2. Tamm Lasers

Plasmonic lasers offer the advantage of mode size below the diffraction limit with lasing under extreme conditions [142, 143]. Unlike conventional lasers with resonant cavity layer, the Tamm plasmon mode itself provides the role of resonant cavity and the control of which leads to easy tailoring of lasing devices and their emission properties in the nanoscale. Symonds *et al*. [144] reported experimental demonstration of lasing in metal dielectric quantum well structure exploiting Tamm plasmons which opens the possibility of integrated micro-laser based devices. Their device was made of a DBR formed by 40 pairs of AlAs/Ga$_{0.95}$Al$_{0.05}$ with a 45 nm silver layer on top for exciting the Tamm mode with lasing wavelength around 857 nm at 77 K. The same group demonstrated lasing in confined Tamm Plasmon structures [145] consisting of metallic microdisks of varying diameters deposited on a bi-dimensional DBR shown in Fig. 11(a). Thirty pairs of AlAs/AlGaAs comprised the active DBR with lasing wavelength at 855 nm at 77K. The emission dispersion of the device with varying pump powers (28 µW to 81 µW) for 4µm diameter Ag microdisks is shown in Fig. 11(b). With the increasing pump power, the strong interaction between confined TPP and exciton modes decreases and the fundamental confined Tamm mode emission becomes more and more intense. At high pump powers the emission is profound in the angular aperture of $14^0$. Fig. 11(c) depicts the increase in emission intensity of the fundamental Tamm mode with varying pump power integrated over the whole numerical aperture. Spectral tuning of the Tamm mode has been obtained by thickness gradient of the layers. The decrease of the lateral size of the structure has led to decrease in lasing threshold. Lheureux *et al*. [146] reported polarization control of the confined Tamm lasers from spatially confined locations. Anisotropic three-dimensional confinement of Tamm modes at the interface of active DBR and silver film facilitated polarized emission with degree of polarization >90%, whereas patterning of top metallic layer with micro-rectangles provided the spatial confinement of lasing. Gubaydullin *et al*. [92] demonstrated enhancement of spontaneous emission in metal/semiconductor Tamm plasmon structures at room temperature. The DBR comprised of 30 pairs of GaAs/Al$_{0.95}$GaAs with three monolayers of InAs as quantum dots for active area with top silver layer to excite the Tamm mode. Toanen *et al*. [41] reported a super Tamm structure based on AlGaAs/AlAs DBR with a SiO$_2$ layer inserted between the DBR and the silver coating with room temperature lasing. Compared to a conventional Tamm laser, the introduction of low index SiO$_2$ layer and adjusting the thickness of last DBR layer help in narrowing spectral emission and decreasing loss that facilitates room temperature lasing. Recently, Xu *et al*. [147] have designed and fabricated a TPP based UV laser using ZnO as an active layer between metal and DBR using the strong field confinement property of Tamm plasmons. This ZnO TPP laser is a cost effective and easy-to-fabricate device with a relatively large sample area. However, challenges remain in Tamm lasers to overcome the losses further, decrease of lasing threshold and room temperature lasing with temporal, spatial, spectral confinement of Tamm lasing mode in 3D for practical, useful laser devices.

### *8. 3. Hot-electron photodetectors*

Traditionally used semiconductor photodetectors are limited by the band gap of the semiconductors. According to Shockley-Queisser limit the conversion efficiency of single junction semiconductors lies approximately to 33%. In recent years, plasmonic structures have been explored for efficient photodetection mechanism [148-150]. Upon photon absorption, the energy is transferred to the electrons in the metal, generates non-thermal electrons with energies well above the Fermi level called as hot electrons [151]. In order to achieve high efficiency hot electron generation and collection, significant light absorption in the metal is necessary through coupling of incident light into surface plasmons. Generally, two structures such as Metal-Insulator-Metal (MIM) structure and Metal-Semiconductor (MS) Schottky junction [152] are used to generate and collect hot carriers in metals. In both cases, the incident light is largely absorbed in one metallic nanostructure contact to excite surface plasmons below the semiconductor band gap. The absorption either leads to direct generation of hot electrons or to surface plasmons. These carriers will diffuse and a fraction of them will pass the dielectric or semiconductor interface. A net current will flow based on the absorption profile within metal layer (layers) and on the voltage established by the energy barrier for hot electrons to travel from one metal to the other metal (semiconductor) [153]. These hot electrons can be used for photodetection in metal/semiconductor or metal/insulator systems [154-156]. SPs generate hot electrons in metals in a sub-wavelength region more effectively due to strong field localization compared to the direct light illumination [157]. Hence using plasmonics, the hot electron photodetection mechanism can be improved to facilitate room temperature below band gap operations [148-150]. But there is a requirement of specially designed sub-wavelength structures for SP based detectors which involves costly and complicated fabrication procedures [158]. Therefore, TPs are used as an alternative easy and planar solution by implying better light trapping abilities, higher absorption, and cost effectiveness. In a TPP structure the metal part is replaced by M-I-M structure for photodetector applications. The hot electrons are produced at the metal-DBR interface through the non-radiative decay of TP resonance. The hot-electrons then reach the metal/semiconductor interface and ultimately collected at the cathode metal layer. It is worth noting that for such systems the electrons flow in both the directions of the metal layer which determines the net photocurrent generated by the device. Hence there is a need of asymmetric absorption in the top and bottom metal layers because more the asymmetry better is the photoresponse.

**Table-1:** Summary of Tamm plasmon based hot electron photodetectors

| Configuration | Spectral range | Peak photoresponsivity |
|---|---|---|
| Silica/Au/Zno/Au/(TiO$_2$/Al$_2$O$_3$)$^8$ [159] | 700-900 nm | 13.7 nA/mW |
| Substrate/(Ge/SiO$_2$)$^3$/Ge/SiO$_2$/ITO-ZnO-Ti/Au [160] | NIR | 8.26 nA/mW |
| Substrate/(TiO$_2$/Al$_2$O$_3$)$^7$/n-Si/TiN [161] | NIR | 26 mA/W |
| Substrate/(Si/SiO$_2$)$^6$/TiO$_2$/Au/ TiO$_2$/Au [162] | NIR | 2 mA/W |
| 1DPC/Au/TiO2/1DPC [163] | 700-1100 | 21.87 mA/W |
| Au/Si/Au/(TiO$_2$/SiO$_2$)$^6$ [164] | 1000-1400 | 16 mA/W |
| Silica/Au/(MoS$_2$/SiO$_2$)$^4$ [165] | 700-1100 | 12.1 mA/W |
| Substrate/Au/(Si/SiO$_2$)$^7$ [166] | NIR | ---- |

Zhang et al. [159] have theoretically compared the grating coupled SP hot electron photodetector to a TP device consisting of a DBR above Au-ZnO-Au structure and found that the photo-responsivity of the TP photodetector is twice (13.7 nA/mW) than the SP based photodetector(6.5 A/mW). In its first experimental observation, a TP-based wavelength selective hot electron photodetector was prepared by depositing a metal-semiconductor-ITO layer over a DBR [160]. The structure consists of Au-(Ti-ZnO)-ITO structure over a 7 layer DBR shown in Fig. 11(d). The device shows a maximum photoresponse of 8.26 nA/mW when the Tamm mode is excited at 1581 nm for the structure. The electrical photo response at zero bias of the device plotted in in Fig. 11(e) shows the wavelength selectivity of the device. It can be seen that the photoresponse decreases by 80% by changing the wavelength of illumination from 1581 nm to 1529 nm. The absorption difference between the Au and ITO layer plotted in Fig. 11(e) follows the same wavelength dependence as the photo-responsivity which means that the photon absorption is proportional to the photocurrent realizing a wavelength selective photodetection. The time dependent photoresponse of the device is shown in Fig. 11(f) for the wavelengths of 1529, 1555, and 1581 nm for an exposure time of 10 s. It clearly shows that maximum photocurrent is generated when the Tamm mode is excited. Wang et al. [161] proposed a more sensitive broad band photodetector of photocurrent up to 26 mA/W at 1140 nm by using TiN as metallic layer in a photonic hetero-structure containing a n-doped Si as spacer layer and DBR made up of 7 pairs of $Al_2O_3$ and $TiO_2$. Shao *et al.* [162] proposed a dual cavity based hot electron photodetector. This device provides double absorption efficiency compared to the single cavity based devices with tripled responsivity of nearly 2 mA/W at 950 nm. Recently, Liang *et al.* [163] have designed ultra-narrowband hot electron photodetector based on coupled dual Tamm plasmons in a Glass/$(SiO_2/TiO_2)^{20}$/Au/$TiO_2$/Au/$(TiO_2/SiO_2)^{13}$ structure. The coupled dual TPs based photodetector with four Schottky junctions in parallel configurations shows responsivity of 21.87 mA/W at the wavelength of 800 nm, which is two times more than that of the conventional devices with two Schottky junctions in series configuration. Apart from these, many more works are reported on TP based photodetectors available in literature [164-167], whose photoresponsivity are listed in Table-1.

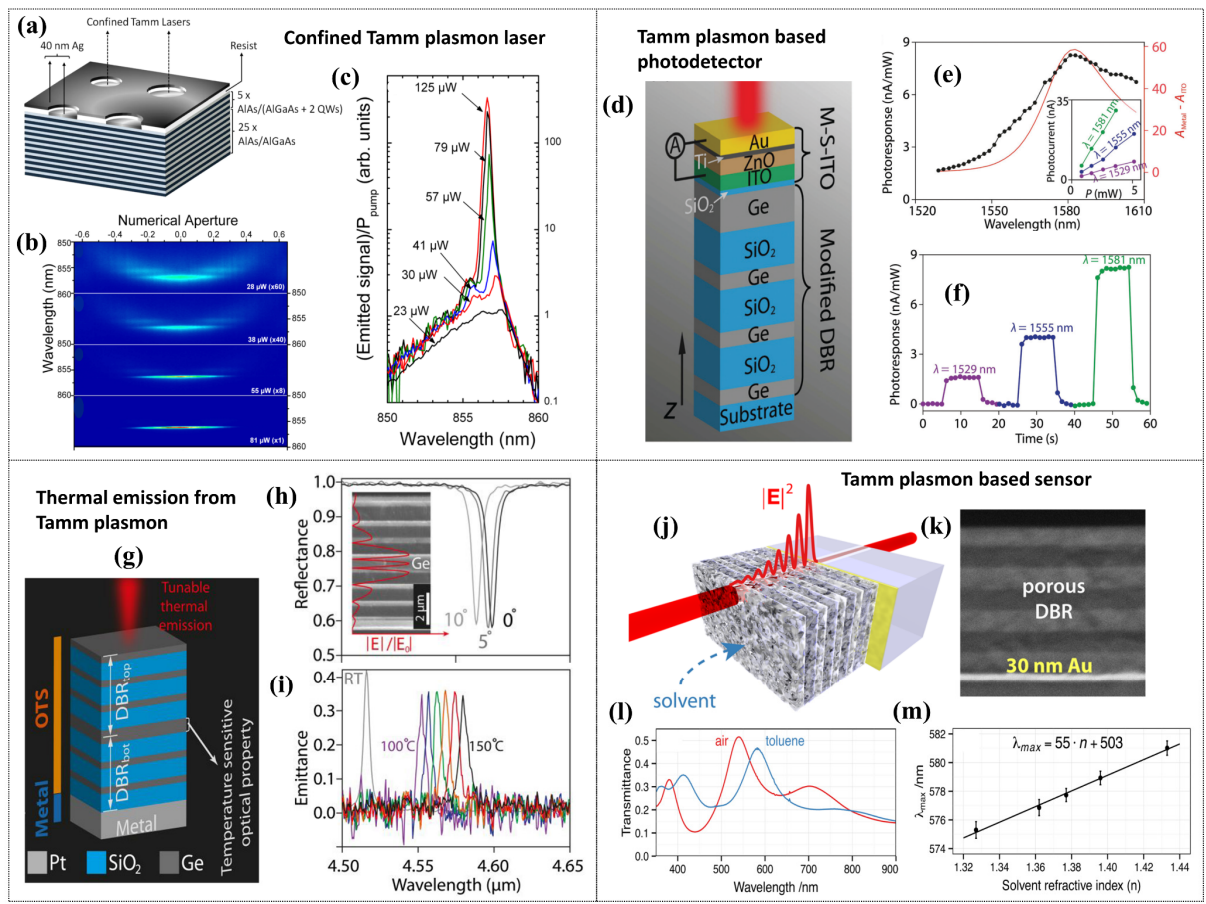

**Fig. 11:** (a) Schematic of a structure for confined Tamm plasmon laser containing five upper AlGaAs layers as active regions, (b) emission contour as a function of wavelength and numerical aperture, and (c) normalized emission spectra for different pump powers [145]. (d) Schematic of a TPP based hot-electron photodetector, (e) photoresponse spectrum of the photodetector with inset plot showing its variation with power of light, and (f) time-dependent photoresponse of the hot-electron photodetector without bias [160]. (g) Schematic of TPP based thermal emitter, (h) reflectance spectra of the emitter at different angle of incidence with inset plot showing the cross-sectional image of the emitter along with the electric field distribution at 0°, and (i) emissivity spectra of the thermal emitter device at different temperatures [168]. (j) Schematic of a mesoporous multilayer Tamm plasmon refractive index sensor and its (k) cross-sectional SEM image showing the multilayer, (l) measured transmittance

30 | P a g e

spectra of the sensor exposed to air (red line) and after immersion in toluene (blue line), and (m) TP mode position as a function of solvent refractive index [169]. All the figures are adapted with permission.

## 8. 4. Thermal emitters

Various narrowband thermal emitters have been demonstrated based on the designs of photonic crystals, surface plasmon polaritons, gap-plasmon resonances, and Tamm plasmon polaritons [170]. Among these, the TPP can be thermally excited and out-coupled from a TP device, producing strong thermal radiation at the resonance wavelength, without the need for complicated nanostructures on the device. Moreover, SPR and gap-plasmon resonance based emitters are associated with certain draw backs with intensity and sharpness of resonance [170-172]. TPs facilitate narrow band wavelength selective thermal emitters which are very useful in sensing applications [173]. The emission from conventional TP structures can be both from metal and DBR sides. But, the emission from the DBR side is more profound compared to the emission from the metal side in terms of emission peak, Q factor and background emission [173]. However, in recent studies more efficient emitters have been designed by considering the coupled Tamm plasmon devices. A narrow band mid IR thermal emitter has been studied by using coupling of micro-cavity mode in the DBR to the TP mode. The hybrid mode was found to show sharper emission band than the conventional TP structure [74]. Fig. 11(g) shows the schematic of a dual DBR ultra-narrow band tunable thermal emitter in the IR regime. The reflection and emission spectra of the hybrid TP device is displayed in Fig. 11 (h) and (i), respectively. It has been observed that the hybrid structure performs better compared to the conventional TP structures. Recently, aperiodic DBR based TP emitters have been studied for suppression of unwanted emission modes and ultra-high Q factors [174]. These devices also provide an extra degree of freedom in terms of spatial control. Wang *et al*. [175] have proposed a modified TP structure in which the last DBR-layer has three times more thickness than in a standard TP structure. The measured emission peak value of the modified TP device at 150ºC is 0.94 with a background of 0.01 and a Q-factor of 48 at a wavelength of 5 μm. The Q-factor of the device is twice as high as that of a standard TP structure. The emission wavelengths in the modified TP device has been tuned from 4.4-5.7 μm by adjusting the last DBR layer thickness, thus demonstrating a fine selection of the emission peak wavelength. An efficient TP based emitter has been designed and experimentally verified containing an aperiodic DBR and n-type doped cadmium oxide (CdO) as metallic layer where the DBR is designed by inverse algorithm based on stochastic gradient descent method [176]. This device is capable of exciting single and multiple resonances suitable for space communication and gas sensors.

## 8. 5. Sensors

### 8. 5. 1. Temperature sensing

Kumar *et al*. [177] have reported a temperature sensor based on Tamm plasmon mode. Their DBR consisted of stratified layers of $Ta_2O_5$ and $SiO_2$ with thin silver film on top. Change in reflectivity due to the Tamm mode excitation was measured as a function of temperature and a sensitivity of $7.8 \times 10^{-4}/^0C$ is estimated for the device within a temperature range of 35-185 $^0C$. Tsurimaki *et al*. [178] have exploited the singularity in phase at near-perfect absorption in a Tamm plasmon structure as a temperature sensor. The structure comprised of alternate layers of $SiO_2/Si_3N_4$ on top of gold film deposited on silicon substrate. This structure exhibits near perfect absorption at 751 nm with reflectivity of $1 \times 10^{-5}$ for p-polarized waves. Spectroscopic ellipsometric measurements were carried out at different temperatures to establish that phase-singularity based measurement being the most sensitive to temperature variations compared to amplitude or reflectance peak shift methods. Maji *et al*. [179] have theoretically demonstrated Tamm plasmon based temperature sensors with sensitivity of $4.5 \times 10^{-4}/^oC$ and $7.5 \times 10^{-4}/^oC$ for TE and TM polarised light, respectively at angle of incidence of 50º. Ahmed *et al*. [180] proposed a high performance temperature sensor exploiting pyroelectric effect and Tamm modes. Their system consists of prism/Ag/($LiNbO_3/SiO_2$) stack with Ag layer exciting the Tamm resonances through the 1D photonic crystal. Thermal characteristics of $LiNbO_3$, determines the position of these resonances depending on the temperature measurable in the range 300-700 K. Ultra-high temperature sensitivity of 1.1 nm/K has been demonstrated.

### 8. 5. 2. Refractive index sensing

Zhang *et al*. [181] have proposed a novel concept of refractive index sensing based on Tamm plasmons which enables to realize sensing over a wide measuring range with high sensitivity. Huang *et al*. [182] have proposed phase detection in spectroscopic ellipsometry for sensitive detection of refractive index, and estimated sensitivity of $2 \times 10^5$/refractive index unit (RIU) for their optimized TP device. Kumar *et al*. [183] proposed hybrid self-referenced refractive index sensor. The sample (analyte) was sandwiched between two metal-DBRs with silver as plasmon-active metal and 8 bilayers of $Ta_2O_5/SiO_2$ formed the DBR. The reflected spectrum of the structure shows two hybrid modes with symmetric and anti-symmetric field distributions about the center with the low-frequency symmetric mode being sensitive to changes in analyte layer. The anti-symmetric mode was used for self-referencing as it is insensitive to changes in analyte layer. Refractive index sensitivity in the range of 65 nm/RIU to 180nm/RIU has been demonstrated. It is expected that sharper TPP mode resonances may lead to better figure of merit of these sensors compared to conventional interferometric or SPR based sensors. In a similar study, $TiO_2/Al_2O_3$ based planar multilayer DBR with gold layer on the substrate to excite Tamm modes was used to demonstrate sensitive analyte detection [184]. Sensitivity of 860 nm/RIU and a figure of merit (FOM) up to 391 was reported by optimal tuning of the Tamm mode. Zhang *et al*. [185] have proposed a fiber based refractive index sensor that exploits hybrid mode of TPP and SPP. Their system consists of a fiber core coated with 1D PC multilayers (pairs of $TiO_2/SiO_2$) with silver on outer layers. Sensitivity in the range of 1310-1420 nm/RIU has been estimated with refractive index in the range 1.33-1.45 and figure of merit of the sensor in the range 62-168 /RIU. Du *et al*. [186] have proposed electromagnetically induced transparency (EIT) like effect in Tamm multilayer structure. They introduced a defect layer in the $TiO_2/SiO_2$ DBR structure which resulted in a sharp peak in transmission, akin to EIT like effect. Ultra-sensitive performance of the sensor with 416 nm/RIU sensitivity and a figure of merit of 682/RIU has been reported. Zaky *et al*. [187] reported an ultra-sensitive gas sensor based on Tamm modes in the IR range. Their structure has comprised of gas cavity sandwiched 1D porous silicon photonic crystal and a silver coated prism. The optimized structure has achieved high sensitivity of $1.9 \times 10^5$ nm/RIU and a low detection limit of $1.4 \times 10^{-7}$ RIU. Keshavarz and Alighanbari [188] have proposed refractive index sensor for terahertz domain using Tamm modes in a DBR with graphene layer. A sensitivity of 0.744 THz/RIU or equivalently, 175.5 μm/RIU, and a FOM of 10.33 /RIU was estimated at 1.132 THz.

### *8. 5. 3. Chemical and bio analyte sensing*

Auguié *et al*. [169] were the early group to establish the application of Tamm modes for sensing applications. The schematic of the sensing structure shown in Fig. 11(j) consists of a mesoporous $TiO_2/SiO_2$ multilayer DBR with Au coating on the substrate for exciting the Tamm plasmon mode. The cross-sectional SEM image of the porous $(TiO_2/SiO_2)^4$ DBR is shown in Fig. 11(k). The dispersion of the target molecules in the porous network of the nanostructured layers changes the effective refractive index of the medium, as a result the reflectivity spectrum of the sensing structure exhibiting TP mode shifts its resonant wavelength position as shown in Fig. 11(l). This shift is used for real time sensitive monitoring. The nanostructure is immersed in a series of alcohols and the shift of the peak is observed. The device shows linear response with changes in refractive index with a sensitivity of 55nm/RIU as shown in Fig. 11(m). The device can be tuned for optimal spectral region of sensing by changing the porosity and thickness of the layers. Li *et al*. [189] reported a magneto-optic optical Tamm plasmon sensor comprising Ce-doped $Y_3Fe_5O_{12}$ (CeYIG) thin film with silver layer deposited on a prism. Layers of the sensing medium and Si constituted the DBR with CeYIG film excited the magneto-optic Tamm mode. It was estimated that a figure of merit (FOM) of 1224.21/RIU for refractive index variation of gas from 1.0000 to 1.0006 can be obtained at 1064 nm excitation. Juneau-Fecteau *et al*. [190] have reported a porous silicon based Tamm sensor and demonstrated shift in resonance peaks with different concentrations of toluene/ethanol solutions. The novelty of their sensor is unlike different high-low index coatings used conventionally in DBR, crystalline silicon was periodically electrochemically anodized and the resultant porous silicon structure was transferred on a gold coated glass substrate. Sensitivity of 139 nm/RIU and a FOM of 4 was demonstrated using this porous silicon Tamm sensor. Zhang *et al*. [90] demonstrated fluorescence emission of TPP coupled nanoholes. The DBR comprised of $Si_3N_4/SiO_2$ layers on glass with Ag coating on top. Experimental demonstration of TPP coupled fluorescence emission of dye filled nano-apertures was demonstrated paving way for sensitive detection of bio-analytes. Balevičius [191] used total internal reflection ellipsometry (TIRE) for analysis of angular spectra of hybrid Tamm-surface plasmon modes and compared with the conventional SPR modes. It was concluded that p-polarized detection of hybrid plasmonic modes lead to enhanced detection sensitivity of the water-ethanol mixture solution. In a recent study, TIRE was used to study protein layer formation exploiting hybrid TPP-SPP interactions [192]. A summary of different TPP based sensors is given in Table-2 indicating the spectral range of applications and respective sensitivity.

**Table-2:** Summary of various sensors reported based on Tamm plasmon mode

| Type | Sensor configuration | Spectral range | Sensitivity | Figure of merit (FOM) |
|---|---|---|---|---|
| **Thermal sensor** | $BK7/(Ta_2O_5/SiO_2)^{10}/Ag$ [177] | 520-650 nm | $\frac{\Delta R_{min}}{\Delta T} = 7.8 \times 10^{-4}$ /°C | --------- |
| | Substrate/Au/$(Si_3N_4/SiO_2)^7$ [178] | 700-780 nm | $\frac{d\lambda}{dT} = 0.0057$ nm/°C | $3.8 \times 10^{-4}$ /°C |
| | | | $\frac{d\psi}{dT} = 0.12$ deg/°C | 0.018 deg/(°C·nm) |
| | Prism/Ag/$(LiNbO_3/SiO_2)^5$ [180] | NIR | $\frac{d\lambda}{dT} = 1.10$ nm/K | 0.012874 / K |
| **Refractive index sensor** | Glass/$(SiO_2/TiO_2)^8$/Au [182] | 600-900 nm | $\frac{d\lambda}{dn} = 2 \times 10^{-5}$ °/RIU | ---------- |
| | $(Si/Air)^4$/Ag [181] | Visible | $\frac{dn}{d\lambda} = 0.012$ RI/nm | ---------- |
| | Substrate/$(SiO_2/Ta_2O_5)^8$/Ag/ Analyte/Ag/$(SiO_2/Ta_2O_5)^8$/ Substrate [193] | Visible | $\frac{d\lambda}{dn} = 65\text{-}180$ nm/RIU | 11-21 /RIU |
| | $SiO_2$ substrate/Au/Analyte/ $(TiO_2/Al_2O_3)^7$ [184] | IR | $\frac{d\lambda}{dn} = 860$ nm/RIU | 391 /RIU |
| | Fused silica fibre core/$(SiO_2/TiO_2)^4$/Ag [185] | NIR | $\frac{d\lambda}{dn} = 550\text{-}1380$ nm/RIU | 62-136 /RIU |

|  | Ag/(TiO$_2$/SiO$_2$)$^5$/Defect layer /(TiO$_2$/SiO$_2$)$^5$ [186] | NIR | $\frac{d\lambda}{dn} = 416$ nm/RIU | 682 /RIU |
|  | Prism/Ag/Gas/porous(Si$_1$/Si$_2$)$^8$/Si [187] | NIR | $\frac{d\lambda}{dn} = 1.9 \times 10^5$ nm/RIU | $3.6 \times 10^5$ /RIU |
|  | Graphene/DBR [188] | IR | $\frac{d\lambda}{dn} = 175.5$ μm/RIU | 10.33 /RIU |
| **Bio-sensor** | Glass/Au/porous(TiO$_2$/SiO$_2$)$^3$/TiO$_2$ [169] | Visible | $\frac{d\lambda}{dn} = 55$ nm/RIU | -------- |
|  | Ag/CeYIG/DBR [189] | NIR | ------------ | 1224.21 /RIU |
|  | Glass/Au/porous Si PC [190] | Visible-NIR | $\frac{d\lambda}{dn} = 139$ nm/RIU | 4 /RIU |
|  | Prism/(TiO$_2$/SiO$_2$)$^6$/Au/BSA or GCSF receptor [192] | Visible-NIR | $\frac{\delta\Lambda}{\delta\lambda} = 53.9$ °/nm | -------- |

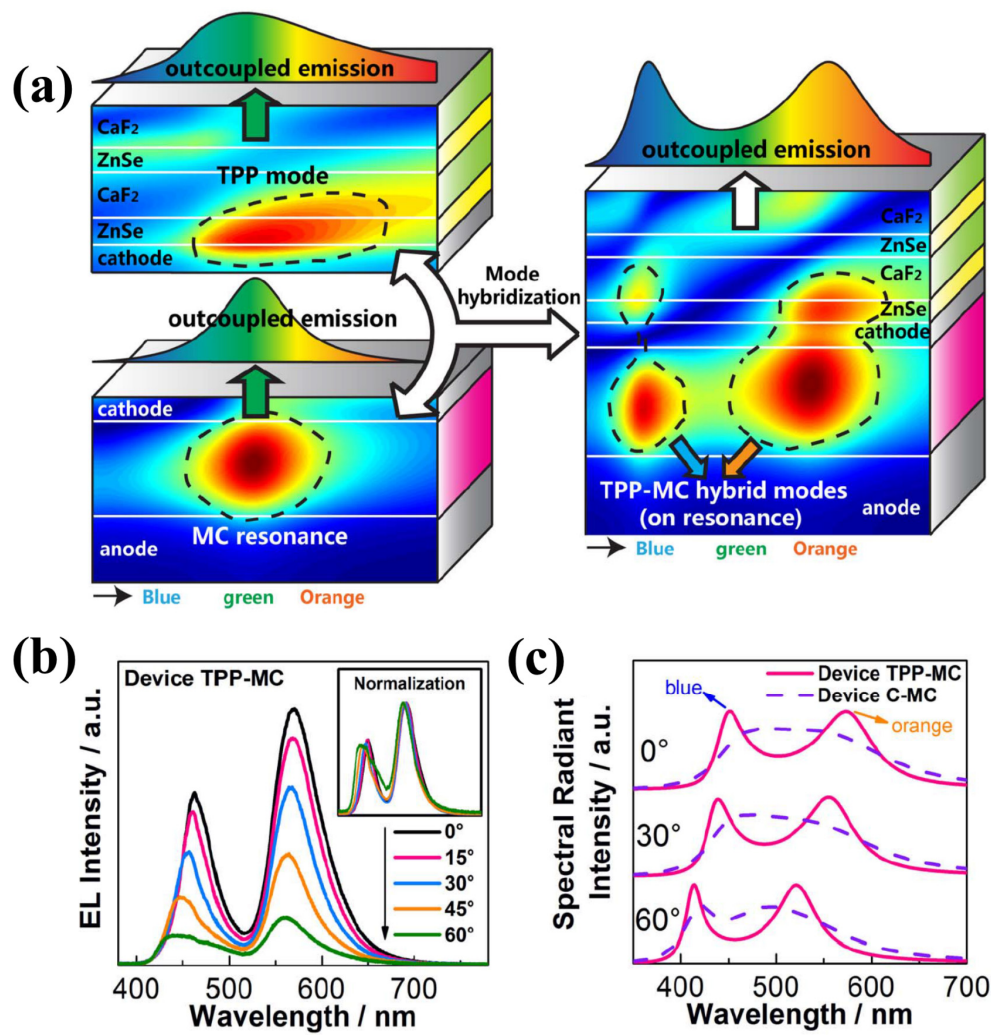

**Fig. 12:** (a) Schematic of the structures exhibiting Tamm plasmon polariton (TPP) mode, micro-cavity (MC) resonance mode, and their coupled hybrid TPP-MC modes used for developing white top-emitting organic light-emitting devices, (b) electroluminescence spectra of the TPP-MC device, and (c) simulated spectral radiant intensity for the TPP-MC device and conventional microcavity (C-MC) device with different viewing angles of 0°, 30°, and 60° [194]. All the figures are adapted with permission.

### *8.6. Other applications*

In addition to the above mentioned applications, novel photonic structures are being designed for exploiting Tamm plasmons for various new applications in active & passive optoelectronics [195-197], resonant cavity LEDs [194], solar thermophotovoltaics [198], and organic solar cell [199-201]. For example, a simple and effective strategy using hybrid TPP-MC modes as shown in Fig. 12 (a) has been executed for improving in both viewing characteristics and light couplings for white top-emitting organic light-emitting device (WTOLED) [194]. The structure shown in left top of this figure exhibits bare TPP mode, while the left bottom structure shows intrinsic MC resonance. The hybrid structure shown in the right side of the Fig. 12(a) generates two TPP-MC hybrid modes under the on-resonance condition *i.e.* with zero detuning wavelength. The light outcoupling efficiency of both the modes are equivalent. The wavelength band corresponding to the resonant modes match fairly well with the blue and orange emission regions. As a result, two complementary-colour-based WTOLEDs has been realized with improved viewing characteristics and electroluminescence (EL) efficiency. The viewing characteristics of the device has been examined by measuring the angular EL spectra of the TPP-MC coupled device as shown in Fig. 12 (b). It can be seen that the coupled modes in the TPP-MC device undergoes slight blue shift with increase in angle of viewing where as their intensity ratio shown in the inset plot of Fig. 12 (b) remains almost constant. The difference between the conventional microcavity (C-MC) and hybrid mode (TPP-MC) based devices can be clearly revealed from the calculated spectral radial intensity for the two devices at different viewing angles as shown in Fig. 12(c). The MC resonance in the C-MC device gets broadened and weakened with blue shift, while the two modes in the TPP-MC device are less affected with slight blue shift, and exhibit a nearly fixed ratio between the intensities of blue and orange peaks. It may be noted that the EL peak shifting range with increasing viewing angle is smaller as compared to that of the calculated spectral radiant intensity (outcoupling efficiency) in OLEDS. It happens due to the invariant intrinsic emission spectrum from the emissive layer. From this study it is evident that the mode hybridization strategy in the proposed WTOLED leads to comparable viewing characteristics like the ITO-based WOLED.

Apart from this, enhanced field amplitude of Tamm plasmon at the metal/photonic crustal interface enables the researchers for the observation of non-linear optical effects. Lee *et al*. [202] have shown extremely large field enhancement factor as large as 3000 at the metal/photonic crystal interface resulting optical bistability at much lower light power compared to the conventional SPP structures. Such structures can be useful for generating strong surface enhanced Raman scattering signals and developing low threshold light power operated optical switching devices. Afinogenov *et al.* [203] have experimentally observed a TPP induced 170-fold enhancement in second harmonic generation (SHG) intensity in comparison with the SHG from the gold film. The SHG enhancement factor exhibits strong angular and polarisation dependence. Yuan *et al.* [204] have reported an ultralow threshold of ~44 kW/cm$^2$ optical bistability in a metal/Kerr nonlinear media which is resulted due to the field enhancement through the resonant transmission in the random layer media and TPPs. Vijisha *et al.* [205] have experimentally demonstrated the enhancement in the nonlinear optical absorption by a factor of nearly 6-fold and optical limiting properties of ZnP+ by using TPP formed at the interface between a thin gold film and a truncated all-polymer Bragg mirror. It is reported that the efficiency of the SHG can be enhanced by two orders of magnitude as compared with a bare metal film, while the overall intensity of the third-harmonic generation can be enhanced by almost five orders of magnitude in the phase matching conditions in the presence of TPPs [206]. These reports suggest that various nonlinear effects can be realized due to TPP induced strong field confinement.

## 9. Conclusions

In summary, this report gives a compact review of properties and applications Tamm plasmon polaritons. They possess polarization dependent parabolic dispersions lying inside the light cone. Apart from planar structure, nano-patterned metal surfaces are observed to generate 3D confined Tamm plasmons. Recent studies have indicated the existence of TP modes even with non-noble metals. However, more research is needed to design and establish a stable TPP structure. A huge scope for further research also exist in metamaterial based TPP devices which can generate Tamm plasmon topological super- lattice. It has been observed that the patterned Tamm device can offer better tunability and higher Q factor than the planar structure. Actively tunable TP devices have been demonstrated which can be tuned by electric field, temperature and refractive index. Liquid crystals have emerged as one of the convenient options to design actively tunable Tamm devices. However, more experimental work needs to be done for demonstrating such devices. We have also discussed the control of spontaneous emission assisted by TPs. TPPs are considered to be suitable for compact room-temperature polariton lasers. A variety of laser systems have been discussed. The coupling of Tamm plasmons to various fundamental modes have also been explained in this article. The hybrid modes are found to depend on geometrical parameters, angle of incidence and polarization. There are reports of using graphene layers to generate coupled modes in hybrid photonic structures. The coupling of TPs with various surface modes like SPP, microcavity mode, Bloch surface waves, and quantum well exciton, offers a vast arena of further research for experimental validation. It has been observed that the addition of monolayers like graphene or transition metal dichalcogenide enhance and impact the performance of TP devices. An overview for different applications of the Tamm plasmon based devices such as lasers, sensors, photodetectors, optical filters, absorbers, and light emitting devices have been presented in this article. In the coming years, the TPP based cost-effective devices will be helpful for the advancement of the energy, medical and industrial markets in numerous applications in our day to day life. In conclusion, it can be said that Tamm plasmon polaritons are an emerging as well as exciting topic in the field of plasmonics based research and device developments.